\documentclass[sigconf]{acmart}
\AtBeginDocument{%
  }

\copyrightyear{2024}
\acmYear{2024}
\setcopyright{rightsretained}
\acmConference[CCS '24]{Proceedings of the 2024 ACM SIGSAC Conference on Computer and Communications Security}{October 14--18, 2024}{Salt Lake City, UT, USA}
\acmBooktitle{Proceedings of the 2024 ACM SIGSAC Conference on Computer and Communications Security (CCS '24), October 14--18, 2024, Salt Lake City, UT, USA}
\acmDOI{10.1145/3658644.3670317}
\acmISBN{979-8-4007-0636-3/24/10}

\usepackage{tikz}
\usepackage{amsmath}
\usepackage{multirow}
\usepackage{diagbox}
\usepackage{graphicx}
\usepackage{colortbl}
\usepackage{array}
\usepackage{filecontents}
\usepackage{enumitem}
\usepackage{subcaption}
\usepackage{graphicx}
\usepackage{pifont}
\usepackage{threeparttable}
\usepackage{tabularx}
\usepackage{xcolor}
\usepackage{hyperref}
\newcolumntype{G}{>{\columncolor{gray!30}}c}

\newcommand{\new}[1]{\textcolor{black}{#1}}

\makeatletter
\gdef\@copyrightpermission{
  \begin{minipage}{0.3\columnwidth}
   \href{https://creativecommons.org/licenses/by-sa/4.0/}{\includegraphics[width=0.90\textwidth]{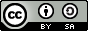}}
  \end{minipage}\hfill
  \begin{minipage}{0.7\columnwidth}
   \href{https://creativecommons.org/licenses/by-sa/4.0/}{This work is licensed under Creative Commons Attribution-ShareAlike 4.0 International.}
  \end{minipage}
  \vspace{5pt}
}
\makeatother

\begin{document}

\title{I Don't Know You, But I Can Catch You: Real-Time Defense against Diverse Adversarial Patches for Object Detectors}


\author{Zijin Lin}
\affiliation{
  \institution{IIE, CAS$^{\dag}$}
  \institution{School of Cyber Security, UCAS$^{\ddag}$}
  \city{Beijing}
  \country{China}
}

\author{Yue Zhao$^*$}
\affiliation{
  \institution{IIE, CAS$^{\dag}$}
  \city{Beijing}
  \country{China}
}

\author{Kai Chen$^*$}
\affiliation{
  \institution{IIE, CAS$^{\dag}$}
  \institution{School of Cyber Security, UCAS$^{\ddag}$}
  \city{Beijing}
  \country{China}
}

\author{Jinwen He}
\affiliation{
  \institution{IIE, CAS$^{\dag}$}
  \institution{School of Cyber Security, UCAS$^{\ddag}$}
  \city{Beijing}
  \country{China}
}

\thanks{$*$ Corresponding author}

\thanks{${\dag}$ Institute of Information Engineering,  Chinese Academy of Sciences}
\thanks{${\ddag}$ University of Chinese Academy of Sciences}

\renewcommand{\shortauthors}{Zijin Lin, Yue Zhao, Kai Chen, \& Jinwen He}

\begin{abstract}
Deep neural networks (DNNs) have revolutionized the field of computer vision like object detection with their unparalleled performance. However, existing research has shown that DNNs are vulnerable to adversarial attacks. In the physical world, an adversary could exploit adversarial patches to implement a Hiding Attack (HA) which patches the target object to make it disappear from the detector, and an Appearing Attack (AA) which fools the detector into misclassifying the patch as a specific object. Recently, many defense methods for detectors have been proposed to mitigate the potential threats of adversarial patches. However, such methods still have limitations in generalization, robustness and efficiency. Most defenses are only effective against the HA, leaving the detector vulnerable to the AA.

In this paper, we propose \textit{NutNet}, an innovative model for detecting adversarial patches, with high generalization, robustness and efficiency. With experiments for six detectors including YOLOv2-v4, SSD, Faster RCNN and DETR on both digital and physical domains, the results show that our proposed method can effectively defend against both the HA and AA, with only 0.4\% sacrifice of the clean performance. We compare NutNet with four baseline defense methods for detectors, and our method exhibits an average defense performance that is over 2.4 times and 4.7 times higher than existing approaches for HA and AA, respectively. In addition, NutNet only increases the inference time by 8\%, which can meet the real-time requirements of the detection systems. Demos and the full paper of NutNet are available at: \url{https://sites.google.com/view/nutnet}.
\end{abstract}

\ccsdesc[500]{Computing methodologies~Object detection}
\ccsdesc[500]{Security and privacy~Software security engineering}

\keywords{Adversarial patch; Physical adversarial defense; Object detectors}


\maketitle

\section{Introduction}
\label{sec:introducion}

Deep neural networks (DNNs) have revolutionized the field of computer vision with their unparalleled performance. Object detection, in particular, has been widely applied in fields with high-security requirements, such as autonomous driving and video surveillance, among others. 
However, existing research~\cite{szegedy2014intriguing,goodfellow2015explaining,carlini2017evaluating,brown2018adversarial} has shown that DNNs are vulnerable to adversarial attacks, which manipulate the input data to produce incorrect results without noticeable changes to human perception. Adversarial attacks come in different forms: adversarial perturbations in the digital world and adversarial patches in the physical world. 
Among these, adversarial patches pose a significant threat to object detector-based applications by making physical adversarial attacks against detectors possible.

The adversary could exploit patches to implement a Hiding Attack (HA) which patches the target object to make it disappear from the detector and an Appearing Attack (AA) which fools the detector into misclassifying the patch as a specific object, leading to potentially catastrophic consequences in the physical world. For example, an attacker could use HA adversarial patches to hide a person in a surveillance video~\cite{Thys_2019_CVPR_Workshops,DBLP:journals/corr/abs-1910-11099,10.1007/978-3-030-58548-8_1,9156452}, leading to a security breach or even a criminal act going undetected. Another potential scenario is that an autonomous car mistaking an AA adversarial patch for a stop sign may stop suddenly and cause accidents~\cite{seeingisnotbelieving}.

Adversarial defenses against adversarial perturbations have been explored extensively, including image scaling, denoising, or compression methods~\cite{gu2015deep,10.1145/3133956.3134057,samangouei2018defensegan,guo2018countering,das2017keeping,dziugaite2016study}. However, these methods are less effective in defending against adversarial patches as patches possess greater robustness. Therefore, recent years have seen the emergence of many defense methods against adversarial patches. Some defenses~\cite{Hayes_2018_CVPR_Workshops,2021arXiv210805075C} utilize model interpretation methods like heat maps or saliency maps to locate and remove the patch region. However, though such defensive methods are effective for classification models, they are not directly applicable to detection models. HA patches and the background show similar effects in interpretable methods against detectors, making it difficult to differentiate them.

Currently, there are also some defense techniques~\cite{Xu_2023_WACV,Liu_2022_CVPR,10.1145/3474085.3475338} proposed for object detection models. These typically depend on an external model for adversarial patch detection, yet these approaches still exhibit limitations in \textit{(L1)} generalization, \textit{(L2)} robustness and \textit{(L3)} efficiency. 
\textit{(L1)} Existing methods have limited generalization. On one hand, most of them focus on detecting HA patches, lacking validity proofs against AA patches. On the other hand, their defense models rely on training datasets containing pre-generated patches and may perform poorly on unseen patches. 
\textit{(L2)} 
These methods may raise concerns about the robustness. Beyond the object detection model, they introduce a defense model to detect or filter adversarial patches, which may also be vulnerable to adversarial examples. Therefore, utilizing adversarial examples to carry out adaptive attacks against both the defense model and object detector raises concerns about the robustness of the defense model.
\textit{(L3)} Some methods present efficiency challenges for object detection systems since their introduced external defense models are not sufficiently lightweight (\emph{e.g.}, segmentation models). However, object detection models require high real-time performance since they are commonly utilized in fields such as autonomous driving or intelligent surveillance.

To improve the above limitations, we propose NutNet, an innovative model for detecting adversarial patches. Regarding \textit{(L1)}, we find that the main cause of the limited generalization in the defense methods is their dependency on features or types of adversarial patches ever learned. To get rid of the dependence on any pre-generated patch, we focus on using only clean samples to train a defense model, without incorporating adversarial patches. In this context, clean samples are treated as in-distribution data for the defense model, while adversarial patches are considered out-of-distribution data. 
Our model is structured as a reconstruction-based autoencoder, leveraging the assumption that in-distribution data can be accurately reconstructed, while out-of-distribution (OOD) samples, such as adversarial patches, cannot. This key distinction allows us to measure the distance between the original input and its reconstructed counterpart, facilitating the detection of OODs, specifically adversarial patches. By compelling the model to recognize clean images, we successfully eliminate the dependency on patch types or features. It's noteworthy that in some cases (\emph{e.g.}, adversarial examples), OODs could also be faithfully reconstructed with autoencoders, which would be further mitigated in our solutions for \textit{(L2)}.

Regarding the robustness of the defense model \textit{(L2)}, we find that though neural networks are always vulnerable to adversarial examples, leveraging such samples to induce a model to produce outputs beyond its capabilities remains challenging. For example, 
if a model is trained to generate cat images and lacks knowledge of unrelated categories like vehicles, it becomes challenging to exploit adversarial examples to deceive it into generating such unrelated categories.
This insight motivates the development of a robust autoencoder, ensuring that the objectives of adversarial attacks fall outside the model's capabilities. To achieve this, we introduce Image-splitting and Destructive Training.

Image-splitting amplifies the distinction between adversarial and normal inputs in feature space. 
In object detection scenarios, adversarial patches typically occupy only a small portion of the image, making them difficult to differentiate. To mitigate this, we divide the image into blocks, treating each as a separate input for the autoencoder. Ideally, when the block size matches the patch size, there should be minimal feature overlap between blocks with and without patches. We determine suitable block sizes considering the distribution of patch sizes observed in previous physical adversarial attacks.
Destructive training further limits the model's capability to reconstruct adversarial patches on the basis of decreased feature overlap between patched and normal images. We do that by deliberately impairing the ability of the autoencoder to generate anything beyond normal samples. 
The decoder, with its limited generative ability, functions like a ``nut'', creating a stable and self-locking system with the encoder (``screw''), hence we name it NutNet. 
Finally, we adhere to the ``less is more'' design principle, introducing minimal model parameters in crafting the autoencoder to address \textit{(L3)}. Our defense model, utilizing only $5k$ parameters, is significantly smaller than the segmentation models used in existing works (1M-parameter for SAC~\cite{Liu_2022_CVPR},  APM~\cite{10.1145/3474085.3475338} based on U-Net, and over 25M-parameter for PatchZero~\cite{Xu_2023_WACV} based on PSPNet).

We conduct extensive experiments in both the digital and physical world to validate the effectiveness of NutNet in defending six detection models, \emph{i.e.}, one-stage detectors (YOLOv2-v4~\cite{redmon2016yolo9000, yolov3, bochkovskiy2020yolov4}, SSD~\cite{liu2016ssd}), two-stage detector (Faster RCNN~\cite{ren2016faster}) and transformer-based detector (DETR~\cite{10.1007/978-3-030-58452-8_13}), on four datasets, \emph{i.e.}, COCO, INRIA, KITTI, and Apricot. The experimental results demonstrate the effectiveness of NutNet in mitigating adversarial patches for both HA and AA. Note that defenses against HA aim to improve the $\text{AP}_{0.5}$ (Average Precision evaluated with an IoU threshold of 0.5) while defenses against AA aim to reduce the $\text{AP}_{0.5}$. Specifically, NutNet reduces $\text{AP}_{0.5}$ of AA from 80.1\% to 2.0\%, and improves $\text{AP}_{0.5}$ of HA from 49.3\% to 72.6\%, outperforming existing works over 4.7 times and 2.4 times, respectively, 
with only a 0.4\% loss of clean data accuracy. In addition, NutNet incurs very little overhead to the model detection, with only an 8\% increase in the inference time, less than most existing defenses we compare. We also verify NutNet with the adaptive attack and find NutNet is still robust due to the contradiction of optimization between bypassing NutNet and achieving the adversarial attack.

\noindent\textbf{Contributions.} The contributions of the paper are as follows:

\begin{itemize}[leftmargin=*,topsep=0pt]
\vspace {3pt}
  \item[$\bullet$] We propose a new defense called NutNet, the first general defense against both HA and AA. By pinpointing distributions of clean samples through a combination of strategies, including a lightweight autoencoder design, patch size level Image-splitting, Destructive Training, and DualMask Generation, NutNet establishes a defense model that is both lightweight and gets improved in performance, generalization, and robustness.
  
\vspace {3pt}
  \item[$\bullet$] We conduct extensive experiments to evaluate NutNet in both the digital and physical worlds. Experimental results demonstrate that NutNet achieves excellent results in defending against HA and AA across different detectors, outperforming existing defense methods over 2.4 times and 4.7 times, respectively.
\end{itemize}

\section{Background and Related Work}
\label{sec:background}

\subsection{Object Detection}
The goal of object detection is to automatically detect and locate objects of interest in an image, and assign each object to one or more predefined classes. Object detectors are mainly classified into two categories, one-stage detectors YOLO~\cite{redmon2016look, redmon2016yolo9000, yolov3, bochkovskiy2020yolov4}, SSD~\cite{liu2016ssd}, RetinaNet~\cite{Lin_2017_ICCV}, DETR~\cite{10.1007/978-3-030-58452-8_13} and two-stage detectors RCNN~\cite{girshick2014rich}, Fast RCNN~\cite{girshick2015fast}, Faster RCNN~\cite{ren2016faster} and Mask RCNN~\cite{he2018mask}, depending on different ways to handle object proposals and object classification. 

One-stage detectors directly predict the class and location of objects in an image, without generating region proposals. They typically use a single network to simultaneously classify objects and predict their bounding boxes. DETR is essentially a one-stage detector that replaces the complex handcrafted operations on detection results, such as non-maximum suppression (NMS), with a transformer-based approach. It transforms the traditional one-stage detection process into an end-to-end framework. While two-stage detectors first generate a set of region proposals in an image, which are then classified as either containing an object or not. These proposals are typically generated by a separate network or module, such as a region proposal network (RPN) in Faster R-CNN. 

\subsection{Adversarial Attacks}
\label{sec:back_attack}
Adversarial attacks involve the intentional introduction of imperceptible perturbations to the input data, causing the model to produce incorrect or unexpected results. The manipulated data is referred to as adversarial examples.
Early studies focused on adversarial examples against image classifiers in the digital world.
\new{Adversarial examples were first discovered by Szegedy \textit{et al.}~\cite{szegedy2014intriguing} and later, Goodfellow \textit{et al.}~\cite{goodfellow2015explaining} proposed the Fast Gradient Sign Method (FGSM) as a fast way to generate adversarial examples. Then new methods like Basic Iterative Method (BIM)~\cite{kurakin2017adversarial}, Project Gradient Descent (PGD)~\cite{madry2017towards} and C\&W attack~\cite{carlini2017evaluating} were proposed to improve FGSM and generate more effective adversarial examples. In addition, researches like~\cite{brown2018adversarial, karmon2018lavan} employed adversarial patches instead of perturbations to execute their attacks, laying the groundwork for adversarial patch attacks against object detectors.}

Recent studies highlight the substantial threat of adversarial attacks in real-world scenarios, especially impacting object detectors. Unlike adversarial perturbations that alter the entire image, real-world adversarial attacks involve attaching an adversarial patch specifically to the target object, allowing modifications only in distinct regions of the image. Adversarial patch attacks fall into two categories: Hiding Attacks (HA) and Appearing Attacks (AA). HAs include targeted HA, where adversarial patches render the detector unable to identify specific objects, and untargeted HA, causing the detector to fail to detect any objects in the image. \new{Note that the misclassification attack is regarded as a special type of targeted HA.} AAs aim to deceive the detector into misinterpreting the patch as a predefined object.
Recent studies~\cite{220580, zolfi2020translucent, Thys_2019_CVPR_Workshops, DBLP:journals/corr/abs-1910-11099, 10.1007/978-3-030-58548-8_1,9156452, Hu_2022_CVPR,lee2019physical,seeingisnotbelieving,10205173,9879058,wang2023does} indicate that HA can involve attaching patches to the camera lens, target object, or picture corner to avoid detection. Zhao \textit{et al.}~\cite{seeingisnotbelieving} demonstrate AA using a printed patch without semantic information, tricking the detector into recognizing it as a stop sign. These attacks highlight the crucial necessity for robust defense mechanisms to guarantee the reliability and safety of real-world detection systems.

\subsection{Adversarial Defense}
\label{sec:back_defense}
Adversarial defenses can be categorized into online defense and offline defense based on their deployment time. Offline defense refers to enhancing the robustness of the model during training, without making any modifications during the prediction stage. Offline defenses typically include adversarial training~\cite{goodfellow2015explaining,tramèr2020ensemble,madry2019deep,wong2020fast}, defensive distillation~\cite{7546524}, etc. 
Online defense, on the other hand, does not modify the model but process images during prediction, using techniques such as denoising~\cite{gu2015deep,10.1145/3133956.3134057,samangouei2018defensegan}, compression~\cite{das2017keeping,dziugaite2016study} and smoothing~\cite{guo2018countering,DBLP:journals/corr/abs-1807-01216}. \new{These defenses are generally effective only against adversarial perturbations. For adversarial patches against classifiers, some explanation-based methods like~\cite{Hayes_2018_CVPR_Workshops,chen2021turning,9283860} use heatmaps or saliency maps to locate the patches and then apply masking or suppression. 
However, the output of a detector is more complex than that of a classifier, and different types of attacks affect the detector's output in various ways. Consequently, it is challenging to directly apply explanation-based methods designed for classifiers to detectors.
Therefore, defenses for adversarial patches on detectors typically rely on sliding masking~\cite{xiang2022objectseeker} or external models~\cite{10.1145/3460120.3484757, Liu_2022_CVPR, Xu_2023_WACV} to mitigate the patches.}

Generally, offline defense is faster but can limited to known threats. For example, a model enhanced with adversarial perturbations may not be able to defend against adversarial patches. The online defense has broader applicability to unknown threats but incurs additional overhead. Below, we explore specific techniques defending against object detectors in each category and delve into their respective limitations.

\vspace{3pt}
\noindent\textbf{Offline Defense. }
In terms of offline defense, Adversarial YOLO~\cite{ji2021adversarial} introduces adversarial patches into the training dataset to enhance the YOLOv2 detector's ability to detect patches. Meanwhile, ROC~\cite{Saha_2020_CVPR_Workshops} fine-tunes the YOLOv2 detector to reduce the reliance on contextual reasoning, thereby minimizing the impact of patches with minimal overlap with the target object. 

\noindent\textit{Limitations. }
Offline defenses involve altering the target model's architecture or parameter weights, potentially affecting its primary task performance. These defenses often only guard against known attack types. For instance, ROC~\cite{Saha_2020_CVPR_Workshops} effectively defends against non-overlapping patches (e.g., those in image corners) but struggles when patches overlap with objects, presenting a notable limitation.

\vspace{3pt}
\noindent\textbf{Online Defense with Sliding Masks.} 
In this defense category, the methods involve analyzing diverse predictions obtained by applying masks to input data. The final prediction is determined by contrasting the model's responses to different masked positions. For instance, PatchCleanser~\cite{279910} fortifies classifiers against adversarial patches, and ObjectSeeker~\cite{xiang2022objectseeker} is tailored for safeguarding object detectors using a similar principle.


\noindent\textit{Limitations.} 
These defense methods often demand creating numerous images with different masks for one input, impacting the model's inference efficiency. In addition, ObjectSeeker~\cite{xiang2022objectseeker} detects anomalies by analyzing diverse detection results from applying various masks to the object detector. However, this approach is limited to defending against HA.

\vspace{3pt}
\noindent\textbf{Online Defense with Texture Smoothing.} 
These methods assume that adversarial patches exhibit higher entropy or more drastic pixel changes than clean images. To counter patches, regions with pronounced texture variations can be detected and smoothed or masked. Jedi~\cite{Tarchoun_2023_CVPR} proposes concealing patches by masking high-entropy image regions. Local Gradient Smooth (LGS)~\cite{DBLP:journals/corr/abs-1807-01216} uses a similar approach, smoothing gradients in high-frequency areas to deactivate patch-like regions, originally designed for classification but adaptable to detection models.

\noindent\textit{Limitations.} 
The assumption underlying these methods is not applicable to adversarial patches with high smoothness. 
Jedi~\cite{Tarchoun_2023_CVPR} relies on two-dimensional entropy for anomaly detection, but this measure primarily reflects pixel change intensity, making it prone to failure with smoother patches or complex backgrounds. LGS~\cite{DBLP:journals/corr/abs-1807-01216} faces a similar challenge, proving ineffective against smoother patches, as some attacks~\cite{Chiang*2020Certified} successfully bypass this method. In Section~\ref{sec:defense digital}, our experiments also show the limited effectiveness of these two defense methods.

\begin{figure*}[ht]
  \centering
  \includegraphics[width=0.75\textwidth]{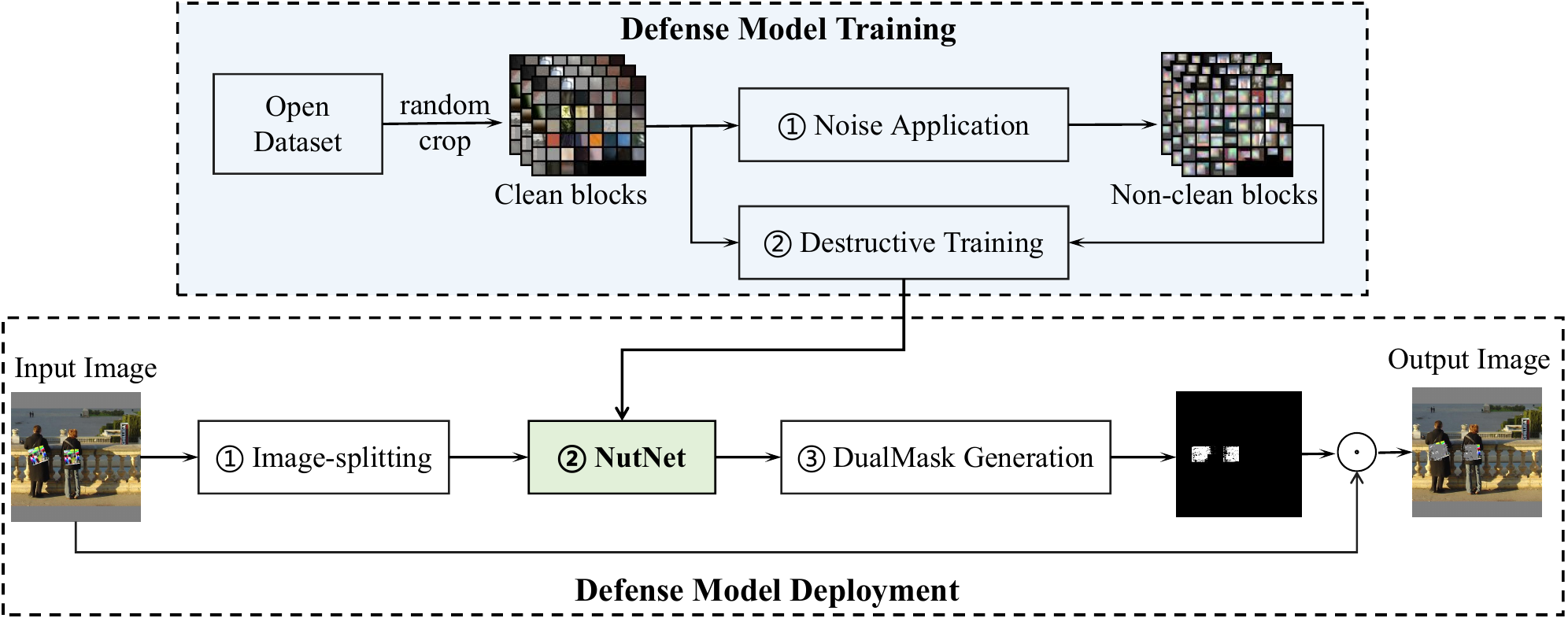}
  \caption{The framework of our defense.}
  \label{fig:framework}
\end{figure*}

\vspace{3pt}
\noindent\textbf{Online Defense with Patch Detection/Segmentation.} 
These methods add an extra model before the object detector to help detect or segment adversarial patches. 
DetectorGuard~\cite{10.1145/3460120.3484757} repurposes a BagNet classifier as an objectness predictor to signal the presence of an adversarial patch but doesn't mitigate or accurately predict it. 
SAC~\cite{Liu_2022_CVPR}, APM~\cite{10.1145/3474085.3475338}, and PatchZero~\cite{Xu_2023_WACV} employ an end-to-end masking strategy, training a patch segmentation model often based on architectures like U-Net~\cite{10.1007/978-3-319-24574-4_28} or PSPNet~\cite{Zhao_2017_CVPR}.

\noindent\textit{Limitations.} 
The training of patch segmentation models, akin to adversarial training, results in a strong coupling with specific downstream tasks. This interdependence poses challenges for transferring these models across different tasks or datasets. 
PatchZero~\cite{Xu_2023_WACV} claims to defend multiple tasks but uses different patch segmentation models for distinct tasks. Diverse feature spaces and vulnerabilities across tasks result in diminished defensive performance when applying a single patch segmentation model to multiple tasks. 
In object detection, challenges emerge with patches targeting diverse object categories. Experiments in Section~\ref{sec:efficiency} demonstrate poor performance when applying a model trained for one patch type to others.

In conclusion, existing defenses against patch attacks on object detectors, including both offline and online methods, exhibit certain limitations. 
Hence, there is an urgent need for comprehensive defense mechanisms that exhibit strong transferability, robustness, low overhead and are capable of effectively guarding against various types of patches.

\section{Overview}
\label{sec:overview}

\subsection{Threat Model}
\label{sec:threat model}
\new{\noindent \textbf{Attacker's goal and capability. }
The attacker aims to fool a real-time object detector by adversarial patches. It is assumed that the attacker has full knowledge of the victim detector, including its parameters and structures, and can generate effective patches using white-box methods. The attacker can manipulate the attack type (HA or AA), size, shape, position, and number of patches used, without prior knowledge by the defenders. Furthermore, the attacker might be aware of the defense deployment including the algorithm and defense model, enabling him to launch adaptive attacks against the defense method.}

\vspace{3pt}
\noindent \new{\textbf{Defender's goal and capability. }
The defender aims to make the detector output the correct prediction without being affected by adversarial attacks. However, the defender is unaware of whether the detector's input is a clean image or an image with an adversarial patch. Such a threat model maximizes the adversary's capabilities and aligns with the real-world scenario faced by the defenders.
The defender employs an online defense approach to process the images instead of employing an offline defense approach to modify the victim model. }

\subsection{Motivation and Defense Intuition}

An effective and real-time defense against adversarial patches is crucial for securing applications reliant on object detection models. While utilizing additional models for adversarial patch detection proves effective, it still faces challenges in generalization, efficiency, and robustness. The use of large detection models often leads to limited efficiency, hindering their application in real-time demanding scenarios such as autonomous driving. Therefore, establishing a small yet effective defense model becomes imperative.

To achieve this, we need to address two questions: how to make the defense model small and how to improve its performance and robustness. 
Intuitively, a model can perform well with few parameters when it doesn't need to learn excessive and complex knowledge. 
As a solution, we deconstruct the intricate functionality of existing defense models, encompassing tasks such as detecting, segmenting, locating, and masking adversarial patches. We preserve only the detection function within the model, employing non-model methods for other functions. This approach leads to a substantial reduction in the model's size, making it lightweight. 
To enhance the lightweight model's ability to detect adversarial patches from input images, we further analyze the differences between adversarial patches and clean images, finding adversarial patches always differ from clean images in semantics, textures, or colors. Therefore, treating adversarial patches as OOD samples for detection, coupled with widening the gap between the distributions of adversarial patches and clean images, provides a potential solution to improve the performance and robustness.

\subsection{Overview of NutNet}
In this paper, we propose NutNet to mitigate the adversarial patch attacks against object detectors. The framework is demonstrated in Figure~\ref{fig:framework}. NutNet, functioning as a reconstruction-based autoencoder, is trained to differentiate between the distribution of clean images and that of non-clean images. In the training of NutNet, we introduce Image-splitting and Destructive Training to enhance the model's ability to reconstruct only the images from a specific clean distribution. By evaluating the reconstruction error, we can effectively separate the clean images from the non-clean images, \emph{i.e.} adversarial patches. Subsequently, we apply the DualMask based on the reconstruction error to remove the identified patch regions, thereby effectively defending against adversarial attacks.

\vspace{3pt}
\noindent \textbf{The defense pipeline.} NutNet contains the following processes 
during the defense phase. Firstly, the input image is split into small non-overlapping blocks, which are subsequently fed into NutNet. NutNet serves as a distribution extractor, which is responsible for reconstructing the blocks. Through an analysis of the reconstruction error, a coarse-grained block-wise mask and a fine-grained pixel-wise mask are generated. These masks are then combined to create a more precise final mask. Subsequently, the patch region is excised from the image based on the final mask, preparing the modified image for detection by the detector.

\section{Approach}
\label{sec:approach}

\subsection{NutNet}
\label{sec:training approach}
We assume that clean images and adversarial patches originate from different distributions. Therefore, we model these two distributions and employ the concept of out-of-distribution detection to identify adversarial patches. Specifically, we can train a distribution extractor to allow only images sampled from the clean distribution to pass through, effectively blocking images from other distributions, thus we can defend against adversarial patches.
We name this distribution extractor NutNet because, like a nut restricting the passage of screws that do not meet its constraints, NutNet imposes strong constraints on its inputs. 

To differentiate the clean sample patterns and adversarial patch patterns based on the data distribution, we first need to establish a model for the data distribution and then measure the distance of a specific sample from this distribution. 
Therefore, we exploit a reconstruction-based autoencoder to learn the representation of the clean data distribution since we have access to the clean training dataset. We achieve this by training it to accurately construct the clean samples but fail to reconstruct non-clean samples effectively. Following this, we view reconstruction error as a form of distance measurement between the input sample and the clean data distribution. Next, we will delve into NutNet from two perspectives: model architecture and training methodology.

\vspace{3pt}
\noindent \textbf{The Architecture.} 
As a distribution extractor, NutNet allows only data from a specific distribution to pass through. Therefore, we design a reconstruction-based autoencoder as the architecture of NutNet, and this architecture directly serves its purpose. Initially, we employ an encoder to reduce the dimensionality of the input image, extracting the underlying distribution. Subsequently, a decoder is used to reconstruct an image from this distribution. If the distribution of the input image aligns with the constraints, the decoder successfully reconstructs it; otherwise, reconstruction cannot take place. 
\new{To reduce additional overhead, we use a lightweight encoder and decoder with only three convolutional or transposed convolutional layers each. This design balances the computational efficiency and the ability to accurately extract the input image's distribution.}

\vspace{3pt}
\noindent \textbf{The Image-splitting Strategy.} 
For NutNet to precisely extract the distribution of input images, we need to constrain its input dimensions. Considering the stealthiness of attacks, the attackers often limit the size of patches. Consequently, in a given input image, an adversarial patch typically occupies only a small fraction. If the autoencoder's input is a complete image, it becomes challenging to meet our requirements for distribution extraction. This is because the majority of regions in such an image belong to the distribution of clean images, which is the distribution we aim to accurately reconstruct. This complexity makes it difficult for a lightweight autoencoder to identify the distribution of small areas containing adversarial patches in the image.

Therefore, we propose the Image-splitting strategy to divide the image into smaller, non-overlapping blocks for a more precise extraction. The Image-splitting addresses the issue mentioned earlier, where adversarial patches occupy only a small fraction. With appropriate block size, some blocks will contain the majority of adversarial patches, while the rest contain almost clean images. This magnifies the contrast between clean and anomalous blocks, making it easier for the autoencoder to learn how to differentiate their distributions. Additionally, compared to processing the entire input image, an autoencoder that handles only small image blocks can have fewer parameters and a faster inference speed. The fewer parameters simplify the autoencoder's feature space, limiting the potential attack surface exposed to adaptive attacks. The faster inference speed minimizes the additional overhead introduced by defense, facilitating real-time detection for the object detector.

Finally, we need to choose an appropriate size for Image-splitting. While adversarial patches cannot be too large for stealthiness, they also cannot be too small for effectiveness. We aim to select an appropriate size that covers the smallest yet effective adversarial patches. In our investigation of existing adversarial attacks, HA patches for hiding a person are usually effective within a range of a few meters, while AA patches for causing a stop sign to appear can be effective from distances exceeding 20 meters. Analyzing the dimensions of the adversarial patches used, we find that the minimum height of effective patches typically needs to be around $1/30$ of the image height. Therefore, we can choose this size for splitting. Taking the input size of $416\times416$ for YOLO detectors as an example, we select a size of $13\times13$ $(416 = 13\times32)$ for splitting the image. In addition, we can also use a size of $26\times26$, or $52\times52$. The smaller the block size, the easier it is for the extractor to filter out patches, but dividing an image into too small blocks may also lead to false positives. The larger the block size, the more features of the image are retained, making it less likely to generate false positives, but the filtering ability for small patches may also decrease. Therefore, we can choose different block sizes for practical applications according to different security requirements. The impact of different block sizes on detection accuracy is evaluated in Section~\ref{sec:abla}.

\vspace{3pt}
\noindent \textbf{The Destructive Training.} 
NutNet is a reconstruction-based autoencoder that can only reconstruct input images conforming to a specific distribution. A straightforward idea is to train it exclusively with data from this particular distribution, which is also employed in some autoencoder-based out-of-distribution (OOD) detection approaches, like MagNet~\cite{meng2017magnet}. However, MagNet can only detect adversarial perturbations against classification models, and cannot meet our requirements.
Training an autoencoder exclusively with data from a specific distribution is impractical for patches. Throughout the training process, the model has only encountered this one distribution, and it learns to reconstruct all input images instead of reconstructing only images from this distribution. In other words, the model establishes a shortcut for an identity transformation from input to output. Therefore, we introduce the Destructive Training, incorporating different distributions during training to destroy the model's reconstruction ability for adversarial patch distributions.

Formally, we use $P_c$ to represent the distribution of clean images (\emph{i.e.}, $x \sim P_c$). 
The autoencoder $\mathcal{E}$ encodes the input images with a convolutional network and then decodes to output the reconstruction of original images with another transposed convolutional network. Let $Dist(\cdot,\cdot)$ be the distance (\emph{i.e.}, reconstruction error) of two images
, and we hope that $\mathcal{E}$ achieves $Dist(x,\mathcal{E}(x)) \ll Dist(x^\prime,\mathcal{E}(x^\prime))$, where $x\sim P_c, x^\prime \nsim P_c$. 
Therefore, the objective function to train $\mathcal{E}$ is as follows:

\begin{equation}
\label{eq:dist}
    \min_{x\sim P_c, x^\prime \nsim P_c} Dist(x,\mathcal{E}(x))-Dist(x^\prime,\mathcal{E}(x^\prime))
\end{equation}

\noindent The first term aims to minimize the gap between the clean image $x$ and its reconstruction $\mathcal{E}(x)$. Conversely, the second term seeks to maximize the gap between the non-clean image $x^\prime$ and its reconstruction $\mathcal{E}(x^\prime)$ (indicated by the negative sign). Optimizing this equation enhances the difference in reconstruction loss between the clean and non-clean images when processed through $\mathcal{E}(x)$, facilitating effective discrimination between them.

Note that we cannot directly sample $x^\prime$ from all distributions that are not $P_c$ in practice and collecting plenty of varied adversarial patterns is costly. Therefore, we simulate this process by adding different noise blocks $t$ to $x\sim P_c$, where $t\sim \mathcal{N}(0,1)$. We scale and stretch $t$ to construct different patterns and paste them onto different regions of $x$. We denote this operation as $H$. \new{Adding noise prevents the reconstruction capability from transferring to non-clean distributions, facilitating filtering out-of-distribution patterns. The impact of the noise distribution will be discussed in Section~\ref{sec:discussion}.}

Then we force the distribution extractor $\mathcal{E}$ to reconstruct these simulated non-$P_c$ distribution images as noise with zero mean and unit variance.
Therefore, Equation~\ref{eq:dist} can be modified as follows, 

\begin{equation}
\begin{aligned}
    \min_{x\sim P_c, t \sim \mathcal{N}(0,1)} &Dist(x,\mathcal{E}(x)) + \mu(\mathcal{E}(H(x,t)) + \\
    &Dist(\sigma^2(\mathcal{E}(H(x,t))),1)
\end{aligned}
\end{equation}

\noindent where $\mu(\cdot)$ represents the mean and $\sigma^2(\cdot)$ represents the variance. The first term comes from Equation~\ref{eq:dist} and helps to control clean performance loss. Concurrently, the last two terms aim to transform non-clean images into noise after passing through $\mathcal{E}$. Through ample sampling, the extractor $\mathcal{E}$ learns the expected appearance of images from distribution $P_c$, accomplishing the extraction of distributions. 

\subsection{The Localization and Mitigation of Patches}
\label{sec:defending approach}
We have previously trained an autoencoder specifically designed for processing small image blocks, and we can easily leverage its characteristics to locate adversarial patches. Specifically, we implement the Image-splitting strategy to divide the image into smaller, non-overlapping blocks, which are subsequently fed into NutNet. Then we calculate the reconstruction error of each block to determine whether an adversarial patch exists. As NutNet is well trained to distinguish the clean image distribution and patch distribution within a small image block, the reconstruction error of clean images is always small while that of patches is always large. As we sequentially reconstruct the image blocks and assemble them into a complete large image, regions with higher reconstruction loss in the resulting image are more likely to contain adversarial patches. In summary, we have achieved the localization of adversarial patches through a straightforward splitting and rearrangement approach.

For images containing adversarial patches, it is crucial to precisely cover the region of the adversarial patch with a mask before inputting it into the object detector to mitigate the impact of the adversarial patches.
While considering the patch localization method discussed earlier, a direct approach is to convert blocks with high reconstruction errors into the mask. However, such a mask is coarse-grained and not accurate enough to be utilized to process the image. In the threat model, we assume that the defender is unaware of the shape of the patch used by the attacker. The coarse-grained mask is composed of square blocks, making it difficult to match the shape of adversarial patches in the image. Additionally, even if the attacker exploits a common square patch, the patch shape may change due to rotation or perspective transformations in the physical world. Processing the input image with a mask that does not match the shape of the patch may disrupt non-patch regions of the image and potentially affect the model's predictions. 

Therefore, we propose DualMask, a method that involves two levels of masking, a coarse-grained one and a fine-grained one, to precisely cover the area of the adversarial patch.
Formally, we initialize a mask $m_1=\{0\}^{H\times W}$ with the same width and height as $x$. Since the input and output of the autoencoder are small blocks of the image, we first calculate the average image reconstruction error before and after the autoencoder for each block $(x)^b_j$ (representing the $j$-th block of image $x$). When the average error $Dist((x)^b_j,\mathcal{E}((x)^b_j))$ exceeds a threshold $\kappa_1$, we set the pixels in the corresponding region in the mask $m_1$ to 1. 
To refine the mask, we conduct fine-grained filtering on the coarse-grained mask. Initially, we set $m_2$ as a zero-filled matrix of dimensions $H\times W$, similar to $m_1$. Then we compute pixel differences across color channels before and after autoencoder processing. When the absolute value of the pixel difference exceeds a threshold $\kappa_2$, we set the corresponding pixel in the mask $m_2$ to 1. 

Once we obtain masks $m_1$ and $m_2$, we form the final mask $m = m_1 \odot m_2$, where $\odot$ denotes element-wise multiplication. It's important to note that utilizing $m_2$ alone, even as a pixel-level mask, is inadequate for image processing without incorporating $m_1$. This precaution is necessary because, even in non-patch regions, some pixels may exhibit notable differences before and after autoencoder processing (\emph{i.e.}, false positives). Relying solely on $m_2$ for image masking might inaccurately filter innocent pixels, disrupting the contextual information of the original image and potentially impacting the model's predictions. Therefore, the inclusion of $m_1$ helps filter out these discrete false positives from clean blocks.
In summary, $m_1$ serves as a coarse-grained mask to identify potential patch locations, while $m_2$ acts as a fine-grained mask within $m_1$. By multiplying $m_1$ and $m_2$, we obtain a mask that better aligns with the shape of the patch. Examples of the DualMask will be shown in Appendix~\ref{sec:appendix dualmask}. 
The related ablation experiments of $m_1$ and $m_2$ could refer to Section~\ref{sec:abla}.

\begin{figure*}[h!]
  \centering
    \begin{subfigure}{0.11\textwidth}
        \centering
        \includegraphics[width=\linewidth]{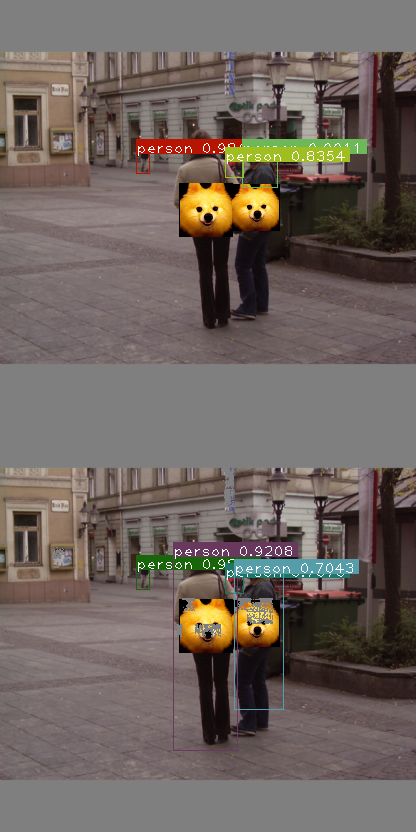}
        \caption{}
        \label{fig:ex_tha1}
    \end{subfigure}
    \begin{subfigure}{0.11\textwidth}
        \centering
        \includegraphics[width=\linewidth]{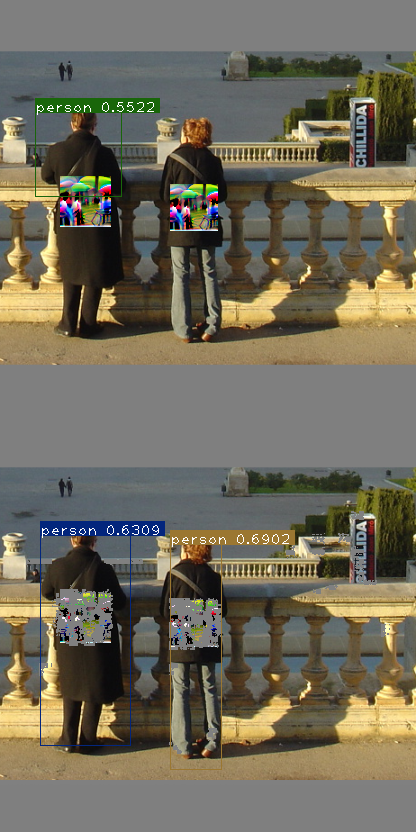}
        \caption{}
        \label{fig:ex_tha2}
    \end{subfigure}
    \begin{subfigure}{0.11\textwidth}
        \centering
        \includegraphics[width=\linewidth]{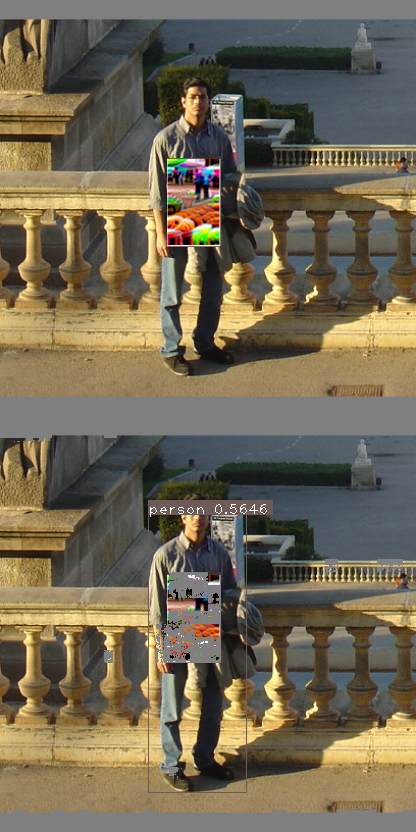}
        \caption{}
        \label{fig:ex_tha3}
    \end{subfigure}
    \begin{subfigure}{0.11\textwidth}
        \centering
        \includegraphics[width=\linewidth]{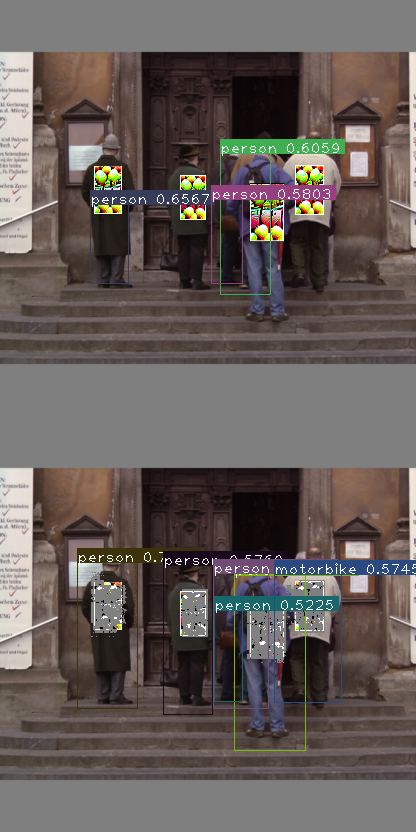}
        \caption{}
        \label{fig:ex_tha4}
    \end{subfigure}
    \begin{subfigure}{0.11\textwidth}
        \centering
        \includegraphics[width=\linewidth]{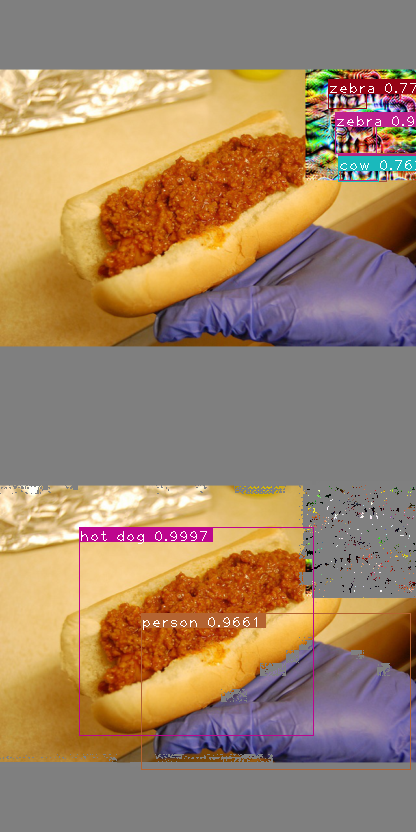}
        \caption{}
        \label{fig:ex_uha1}
    \end{subfigure}
    \begin{subfigure}{0.11\textwidth}
        \centering
        \includegraphics[width=\linewidth]{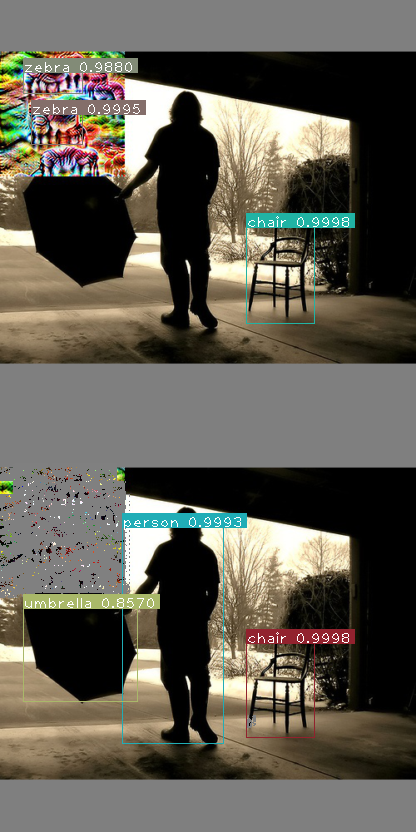}
        \caption{}
        \label{fig:ex_uha2}
    \end{subfigure}
    \begin{subfigure}{0.11\textwidth}
        \centering
        \includegraphics[width=\linewidth]{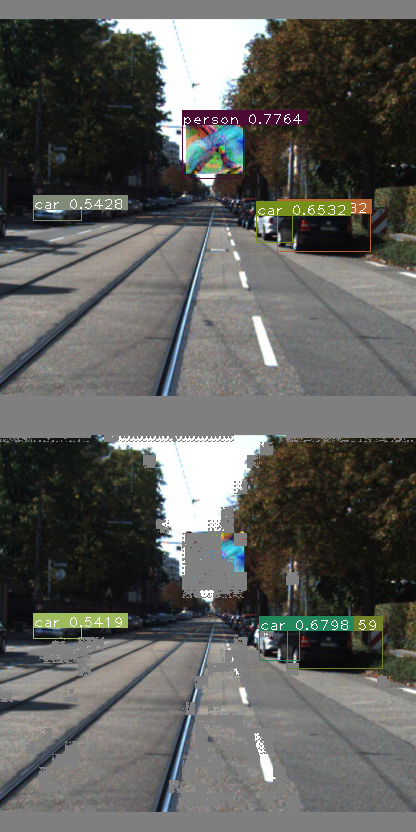}
        \caption{}
        \label{fig:ex_aa1}
    \end{subfigure}
    \begin{subfigure}{0.11\textwidth}
        \centering
        \includegraphics[width=\linewidth]{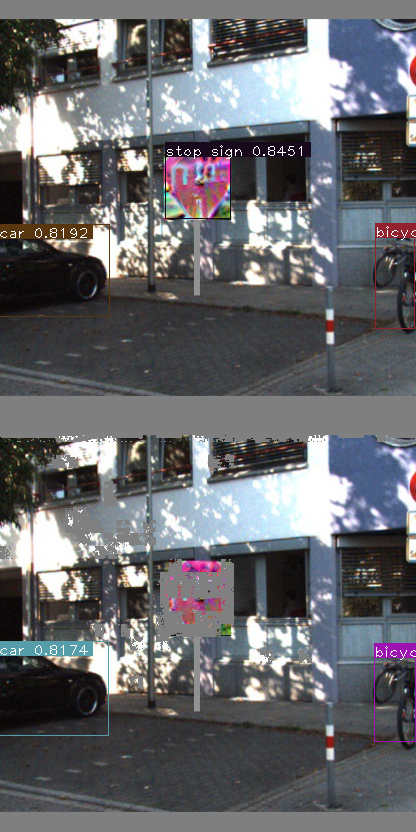}
        \caption{}
        \label{fig:ex_aa2}
    \end{subfigure}
  \caption{Examples of detection results on images with adversarial patches with NutNet. The top images show predictions on the original patched images, while the bottom images show predictions on the patched images with NutNet.}
  \label{fig:digital example}
\end{figure*}

\section{Evaluation}
\label{sec:evaluation}

In this section, we implement our defense with six object detectors and evaluate the efficiency and the defensive performance against different kinds of patches, 
comparing with four existing open-source defense methods~\cite{Liu_2022_CVPR, Tarchoun_2023_CVPR, DBLP:journals/corr/abs-1807-01216, xiang2022objectseeker}.
Furthermore, we conduct an ablation study and analyze the performance of NutNet against adaptive attacks to further investigate its effectiveness.

\subsection{Experimental Setup}
\label{sec:exp setup}


\vspace{3pt}
\noindent\textbf{Object detectors. }
Due to the decoupling between NutNet and the object detectors, it is easy to evaluate NutNet on different object detectors. We select YOLOv2-v4~\cite{redmon2016yolo9000,yolov3,bochkovskiy2020yolov4}, SSD~\cite{liu2016ssd}, Faster RCNN~\cite{ren2016faster} and DETR~\cite{10.1007/978-3-030-58452-8_13} as the victim object detectors, which include traditional convolutional one-stage, two-stage detection architectures and the transformer-based architecture. The backbones of YOLOv2-4 and SSD are Darknet-19, Darknet-53, CSPDarknet-53 and VGG-16 respectively, while the backbones of Faster RCNN and DETR are ResNet-50. SSD is trained on the VOC (PASCAL Visual Object Classes)~\cite{voc} dataset while the other models are trained on COCO (Microsoft Common Objects in Context)~\cite{lin2015microsoft} dataset.

\vspace{3pt}
\noindent\textbf{Attack types. } 
We evaluate the defensive performance of NutNet against the Hiding Attack (HA) and the Appearing Attack (AA), where the HA includes the targeted HA and the untargeted HA. 
The attacks we evaluated are elaborated in Section~\ref{sec:back_attack}.

\vspace{3pt}
\noindent\textbf{Patch sizes. } For adversarial patch attacks in the digital world, we set different sizes for different patch types. 
For the targeted HA patches, we scale each patch to one-fifth of the length of the diagonal of the corresponding bounding box of the object and paste them over target objects, following the setup of~\cite{Thys_2019_CVPR_Workshops}. To enhance the attack success rate of patches with lower efficacy at this size, we further increase their dimensions by 25\%, resulting in an improved attack success rate. Comprehensive details regarding the size of the adversarial patches will be provided in Appendix~\ref{sec:appendix sizes}. 
In the case of untargeted HA patches, we scale each patch following the specifications outlined in~\cite{lee2019physical} and affix them to the corners of the image to minimize interference with other objects.
As for the AA patches, we randomly adjust the size and position them within the image, simulating the capturing of these objects at varying distances. 
For physical-world adversarial patch attacks, we print the patches at a size of 40cm by 40cm, aligning with the setup detailed in~\cite{Thys_2019_CVPR_Workshops}. The patches with natural lookings are printed at a size of 50cm by 50cm, as those patches do not achieve the same level of attack effectiveness as regular patches of identical size.

\vspace{3pt}
\noindent\textbf{Metrics. } 
We use the Average Precision (AP) metric, which is a commonly used metric in object detection evaluation. A higher AP indicates better performance, as it reflects the ability of the detector to accurately locate and classify objects in an image. AP is defined as the average precision values at different recalls. In our evaluation, we compute AP at an IoU (Intersection over Union) threshold of 0.5, denoted as $\text{AP}_{0.5}$. For the targeted HA and the AA, we evaluate the AP for the target class, while for the untargeted HA, we evaluate the mean AP (mAP) across all classes. 
The HA tends to decrease the AP while AA can lead to an increase in AP. Therefore, for HA, a higher AP indicates better defensive performance, while for AA, a lower AP indicates better defensive performance.

\vspace{3pt}
\noindent\textbf{Baseline. } 
Given our evaluation across diverse architectures of object detection models, we opt for online defense solutions renowned for their superior generalization. The comparison with offline defenses is reserved for presentation in Appendix~\ref{sec:appendix offline evaluation}. 
Our benchmark includes the following open-source approaches: \textbf{SAC}~\cite{Liu_2022_CVPR}, \textbf{Jedi}~\cite{Tarchoun_2023_CVPR}, \textbf{LGS}~\cite{DBLP:journals/corr/abs-1807-01216}, and \textbf{OS}~\cite{xiang2022objectseeker}. The choice of these open-source approaches ensures a comprehensive comparison and evaluation across different online defense strategies. 

\vspace{3pt}
\noindent\textbf{NutNet settings. } As is mentioned in Section~\ref{sec:training approach}, we can choose different block sizes according to different requirements. In Section~\ref{sec:defense digital}-~\ref{sec:training approach}, we use the block with a size of $13\times13$, \emph{i.e.}, a $416\times416$-size image will be split into $32\times32$ blocks and then fed into the distribution extractor. We will evaluate the performance of the extractor with other configurations in Section~\ref{sec:abla}. 

\subsection{Effectiveness}
\label{sec:defense digital}
In this section, we evaluate the defensive performance of NutNet and baseline defense methods on six object detectors, including YOLOv2-v4, SSD, Faster RCNN, and DETR, against different types of adversarial patches. 
Examples of defended detection results on images with patches are shown in Figure~\ref{fig:digital example}, where Figure~\ref{fig:ex_tha1}-\ref{fig:ex_tha4} for targeted HA, Figure~\ref{fig:ex_uha1}-\ref{fig:ex_uha2} for untargeted HA and Figure~\ref{fig:ex_aa1}-\ref{fig:ex_aa2} for AA. Besides, we also compare our method with four benchmark defense methods for object detectors. 

\begin{table*}[t]
\caption{Performance ($\text{AP}_{0.5}$) of various defense methods on the INRIA dataset with different patches and models. The gray-background result is from the patch without transformations, and the white-background result is from the patch with transformations (e.g., blurring, rotation, and translation). Bold values indicate the maximum performance.}
\centering
\scriptsize
\begin{threeparttable}
    \begin{tabular}{c|c|Gc|Gc|Gc|Gc|Gc|Gc}
    \hline
      Model & Patch & \multicolumn{2}{c|}{Vanilla} & \multicolumn{2}{c|}{SAC~\cite{Liu_2022_CVPR}} & \multicolumn{2}{c|}{Jedi~\cite{Tarchoun_2023_CVPR}} & \multicolumn{2}{c|}{LGS~\cite{DBLP:journals/corr/abs-1807-01216}} & \multicolumn{2}{c|}{OS~\cite{xiang2022objectseeker}} & \multicolumn{2}{c}{NutNet (ours)} \\
    \hline
      \multirow{9}{*}{YOLOv2}  
        & clean & \multicolumn{2}{c|}{0.847} & \multicolumn{2}{c|}{\textbf{0.885}} & \multicolumn{2}{c|}{0.839} & \multicolumn{2}{c|}{0.874} & \multicolumn{2}{c|}{0.825} & \multicolumn{2}{c}{0.868}      \\
        & AdvPatch (a)~\cite{Thys_2019_CVPR_Workshops} & 0.310 & 0.218 & 0.439 & 0.340 & 0.553 & 0.517 & 0.546 & 0.373 & 0.061 & 0.015 & \textbf{0.780} & \textbf{0.786}  \\
        & AdvPatch (b)~\cite{Thys_2019_CVPR_Workshops} & 0.671 & 0.638 & 0.639 & 0.687 & 0.623 & 0.619 & \textbf{0.778} & 0.755 & 0.565 & 0.549 & 0.771 & \textbf{0.761}  \\
        & AdvPatch (c)~\cite{Thys_2019_CVPR_Workshops} & 0.581 & 0.561 & 0.526 & 0.473 & 0.566 & 0.553 & 0.632 & 0.599 & 0.142 & 0.111 & \textbf{0.772} & \textbf{0.784}  \\
        & AdvPatch (d)~\cite{Thys_2019_CVPR_Workshops} & 0.622 & 0.627 & 0.546 & 0.537 & 0.602 & 0.614 & 0.556 & 0.655 & 0.179 & 0.153 & \textbf{0.788} & \textbf{0.801}  \\
        & AdvT-shirt~\cite{DBLP:journals/corr/abs-1910-11099} & 0.497 & 0.395 & 0.443 & 0.431 & 0.551 & 0.523 & 0.488 & 0.510 & 0.091 & 0.028 & \textbf{0.754} & \textbf{0.724}    \\
        & AdvCloak~\cite{10.1007/978-3-030-58548-8_1}& 0.431 & 0.324 & 0.522 & 0.228 & 0.569 & 0.536 & 0.567 & 0.437 & 0.177 & 0.069 & \textbf{0.760} & \textbf{0.740}  \\
        & NaturalPatch~\cite{Hu_2021_ICCV} & 0.637 & 0.476 & 0.575 & 0.374 & 0.505 & 0.477 & 0.547 & 0.390 & 0.367 & 0.214 & \textbf{0.708} & \textbf{0.649} \\
        & AdvTexture~\cite{Hu_2022_CVPR} & 0.647 & 0.568 & 0.540 & 0.363 & 0.625 & 0.604 & 0.479 & 0.367 & 0.181 & 0.089 & \textbf{0.781} & \textbf{0.765} \\
    \hline
      \multirow{3}{*}{YOLOv3}  
        & clean & \multicolumn{2}{c|}{0.961} & \multicolumn{2}{c|}{0.961} & \multicolumn{2}{c|}{\textbf{0.964}} & \multicolumn{2}{c|}{0.962} & \multicolumn{2}{c|}{0.956} & \multicolumn{2}{c}{0.951}      \\
        & AdvPatch~\cite{Thys_2019_CVPR_Workshops}\tnote{*} & 0.206 & 0.341 & 0.456 & 0.480 & 0.630 & 0.636 & 0.450 & 0.733 & 0.281 & 0.289 & \textbf{0.837} & \textbf{0.823} \\
        & NaturalPatch~\cite{Hu_2021_ICCV} & 0.471 & 0.448 & 0.582 & 0.444 & 0.618 & 0.603 & 0.510 & 0.460 & 0.475 & 0.468 & \textbf{0.783} & \textbf{0.748} \\
    \hline
      \multirow{3}{*}{YOLOv4}  
        & clean & \multicolumn{2}{c|}{\textbf{0.955}} & \multicolumn{2}{c|}{0.948} & \multicolumn{2}{c|}{0.934} & \multicolumn{2}{c|}{0.948} & \multicolumn{2}{c|}{0.954} & \multicolumn{2}{c}{0.948}      \\
        & AdvPatch~\cite{Thys_2019_CVPR_Workshops}\tnote{*} & 0.417 & 0.473 & 0.490 & 0.476 & 0.597 & 0.529 & 0.659 & 0.764 & 0.556 & 0.576 & \textbf{0.829} & \textbf{0.834} \\
        & NaturalPatch~\cite{Hu_2021_ICCV} & 0.877 & 0.810 & 0.723 & 0.676 & 0.728 & 0.734 & 0.879 & 0.836 & \textbf{0.881} & 0.845 & 0.851 & \textbf{0.850} \\
    \hline
      \multirow{2}{*}{SSD}  
        & clean & \multicolumn{2}{c|}{0.855} & \multicolumn{2}{c|}{0.855} & \multicolumn{2}{c|}{0.829} & \multicolumn{2}{c|}{0.834} & \multicolumn{2}{c|}{\textbf{0.887}} & \multicolumn{2}{c}{0.836}      \\
        & AdvPatch~\cite{Thys_2019_CVPR_Workshops}\tnote{*} & 0.412 & 0.308 & 0.435 & 0.440 & 0.520 & 0.489 & 0.654 & 0.581 & 0.318 & 0.225 & \textbf{0.743} & \textbf{0.725}\\
    \hline
      \multirow{3}{*}{FRCNN}  
        & clean & \multicolumn{2}{c|}{\textbf{0.961}} & \multicolumn{2}{c|}{\textbf{0.961}} & \multicolumn{2}{c|}{0.960} & \multicolumn{2}{c|}{0.960} & \multicolumn{2}{c|}{0.411} & \multicolumn{2}{c}{0.956}      \\
        & AdvPatch~\cite{Thys_2019_CVPR_Workshops}\tnote{*} & 0.758 & 0.752 & 0.600 & 0.552 & 0.766 & 0.733 & 0.846 & \textbf{0.849} & 0.245 & 0.247 & \textbf{0.848} & 0.845\\
        & NaturalPatch~\cite{Hu_2021_ICCV} & 0.782 & 0.737 & 0.574 & 0.735 & 0.795 & 0.785 & 0.838 & \textbf{0.804} & 0.248 & 0.228 & \textbf{0.877} & 0.741\\
    \hline
      \multirow{2}{*}{DETR}
      & clean & \multicolumn{2}{c|}{0.857} & \multicolumn{2}{c|}{0.915} & \multicolumn{2}{c|}{0.868} & \multicolumn{2}{c|}{0.919} & \multicolumn{2}{c|}{0.885} & \multicolumn{2}{c}{\textbf{0.924}}      \\
        & AdvPatch~\cite{Thys_2019_CVPR_Workshops}\tnote{*} & 0.769 & 0.764 & 0.771 & 0.774 & 0.732 & 0.687 & 0.801 & 0.793 & 0.716 & 0.704 & \textbf{0.824} & \textbf{0.814} \\
    \hline
    \end{tabular}
\begin{tablenotes}
    \footnotesize
    \item[*] The original AdvPatch was generated based on YOLOv2, and we have expanded its algorithm to accommodate other models. Hereafter we use "*" to represent the patches generated by the original method on other models.
\end{tablenotes}
\end{threeparttable}

\label{tab:tha}
\end{table*}

\vspace{3pt}
\noindent\textbf{Targeted Hiding Attack. }
For targeted HA, we use the INRIA dataset~\cite{maggiori2017dataset}, an object detection dataset with annotations for humans, to evaluate the defensive effectiveness against different kinds of patches. 
For YOLOv2, we use the patches from existing attacks~\cite{Thys_2019_CVPR_Workshops,DBLP:journals/corr/abs-1910-11099,10.1007/978-3-030-58548-8_1,Hu_2021_ICCV, Hu_2022_CVPR}. \new{Specifically,~\cite{Thys_2019_CVPR_Workshops} provides four types of patches: those minimizing only the objectness score (denoted as (a) in Table~\ref{tab:tha}), only the classification score (denoted as (b)), and both (denoted as (c) and (d)). }
For other models with less or even no existing patches, we exploit the code of~\cite{Thys_2019_CVPR_Workshops} to generate patches, as a complementary to existing attacks. 
We test the detection accuracy of each patch with and without applying realistic transformations (such as blur, rotation, translation, etc.) to verify their efficiency and their robustness against transformations.
The experimental results are shown in Table~\ref{tab:tha}, where each patch has two evaluation results: the white-background result is obtained from evaluating the patch without any transformations, while the gray-background result is obtained from evaluating the patch with transformations.


Experimental results show that NutNet can effectively increase the corresponding model's $\text{AP}_{0.5}$ to 0.7 or above when facing the majority of untargeted HA patches with or without transformations, which is close to the detection accuracy of those models on clean images. \new{It is worth noting that among the four adversarial patches used by AdvPatch~\cite{Thys_2019_CVPR_Workshops}, there are patches that either make the object disappear directly or cause the object to be misclassified. However, NutNet can defend against all of them.}
Besides, NutNet only causes a very minor and negligible loss of 0.4\% on clean performance, and it can even improve the clean performance of the YOLOv2 and DETR.

\begin{table*}[th]
\caption{Performance ($\text{mAP}_{0.5}$) of various defense methods under untargeted HA patch against YOLO detectors on the COCO2014 dataset. The values marked in bold are the maximum values.}
\centering
\scriptsize
    \begin{tabular}{c|c|c|c|c|c|c|c}
        \hline
        Model & Patch & Vanilla & SAC~\cite{Liu_2022_CVPR} & Jedi~\cite{Tarchoun_2023_CVPR} & LGS~\cite{DBLP:journals/corr/abs-1807-01216} & OS~\cite{xiang2022objectseeker} & NutNet (ours) \\
        \hline
        \multirow{2}{*}{YOLOv2}  & clean & 0.380 & \textbf{0.387} & 0.211 & 0.362 & 0.312 & 0.377 \\
        & PAPatch~\cite{lee2019physical} & 0.312 & 0.339 & 0.199 & 0.321 & 0.278 & \textbf{0.351} \\
        \hline
        \multirow{2}{*}{YOLOv3}  & clean & 0.513 & \textbf{0.518} & 0.439 & 0.462 & 0.488 & 0.494 \\
        & PAPatch~\cite{lee2019physical} & 0.299 & 0.373 & 0.395 & 0.376 & 0.390 & \textbf{0.438} \\
        \hline
        \multirow{2}{*}{YOLOv4}  & clean & 0.499 & 0.498 & 0.427 & 0.435 & \textbf{0.529} & 0.485 \\
        & PAPatch~\cite{lee2019physical} & 0.425 & 0.445 & 0.400 & 0.412 & \textbf{0.482} & 0.440 \\
        \hline
    \end{tabular}
\label{tab:uha}
\end{table*}

\begin{table*}[th]
\caption{Performance ($\text{AP}_{0.5}$) of various defense methods under AA patches against different models. Note that the stop sign patch and the person patch are all generated based on YOLOv2. The values marked in bold are the minimum values.}
\centering
\scriptsize
    \begin{tabular}{c|c|c|c|c|c|c|c}
        \hline
        Model & Patch & Vanilla & SAC~\cite{Liu_2022_CVPR} & Jedi~\cite{Tarchoun_2023_CVPR} & LGS~\cite{DBLP:journals/corr/abs-1807-01216} & OS~\cite{xiang2022objectseeker} & NutNet (ours) \\
        \hline
        \multirow{2}{*}{YOLOv2}  & stop sign & 0.900 & 0.990 & 0.861 & 0.939 & 0.989 & \textbf{0.041} \\
        & person & 0.972 & 0.968 & 0.906 & 0.926 & 0.940 & \textbf{0.014}\\
        \hline
        \multirow{2}{*}{YOLOv3}  & stop sign & 0.806 & 0.847 & 0.756 & 0.636 & 0.987 & \textbf{0.081}  \\
        & person & 0.952 & 0.930 & 0.892 & 0.727 & 0.938 & \textbf{0.002} \\
        \hline
        \multirow{2}{*}{YOLOv4}  & stop sign & 0.917 & 0.917 & 0.888 & 0.848 & 0.985 & \textbf{0.013}  \\
        & person & 0.885 & 0.885 & 0.807 & 0.496 & 0.901 & \textbf{0.000} \\
        \hline
        SSD  & person & 0.843 & 0.843 & 0.810 & 0.524 & 0.896 & \textbf{0.000}  \\
        \hline
        \multirow{2}{*}{FRCNN}  & stop sign & 0.782 & 0.715 & 0.648 & 0.645 & 0.716 & \textbf{0.001} \\
        & person & 0.277 & 0.277 & 0.161 & 0.252 & 0.035 & \textbf{0.000} \\
        \hline
        \multirow{2}{*}{DETR}  & stop sign & 0.709 & 0.711 & 0.255 & 0.390 & 0.839 & \textbf{0.003} \\
        & person & 0.767 & 0.767 & 0.663 & 0.621 & 0.795 & \textbf{0.074} \\
        \hline
    \end{tabular}
\label{tab:aa}
\end{table*}

As a comparison, four benchmark defense approaches do not significantly impact the detection accuracy of the object detector on clean data. However, their defensive effectiveness is almost universally far inferior to NutNet.
The defensive performance of SAC is not consistent. Some patches make SAC slightly enhance the model's detection accuracy, but there are also cases where SAC results in a decrease in detection accuracy. This instability in performance could be attributed to the fact that the adversarial patches we used may not necessarily be the same as those used during the training of SAC's segmentation model. 
Jedi, when confronted with the NaturalPatch on YOLOv2 and YOLOv4, leads to a decrease in detection accuracy. However, in other scenarios, it generally results in a slight improvement in detection accuracy.
OS, on the other hand, can cause significant accuracy drops in the defense of YOLOv2, SSD and Faster RCNN, which could be due to the models being significantly impacted by the sliding masks and making lots of false-positive predictions. For other models, the impact of using OS for defense on detection accuracy is minimal. 
As for LGS, it may lead to an accuracy decrease when facing natural patches in YOLOv2, but it can provide some defense effectiveness in other cases. However, its improvement in detection accuracy is far less significant compared to NutNet.

\vspace{3pt}
\noindent\textbf{Untargeted Hiding Attack. }
For untargeted HA, we use the COCO2014 dataset, an object detection dataset with annotations for 80 object classes, to evaluate the defense effectiveness against the PAPatch proposed in~\cite{lee2019physical}. 
Note that the PAPatch is generated based on the YOLOv3 detector, and we also conduct transfer attacks using PAPatch against YOLOv2 and YOLOv4.
The patch is placed in the corner of the images in the COCO2014 validation set (40504 images) to minimize the obstruction of objects in the image. The experimental results are shown in Table~\ref{tab:uha}. 

Experimental results indicate that NutNet is effective against PAPatch on all three models, with the most significant increase of over 46.5\% (from 0.299 $\text{mAP}_{0.5}$ to 0.438 $\text{mAP}_{0.5}$) on YOLOv3. As for YOLOv2 and YOLOv4, these two models are less affected by the patch, but NutNet still manages to improve the detection accuracy to some extent. 
As a comparison, 
SAC can slightly enhance the model's detection accuracy on clean images, but the improvement is also relatively limited when facing adversarial patches. 
Jedi provides defense only against PAPatch for YOLOv3. In other cases, it results in a decrease in detection accuracy, including on clean images. This indicates that its assumption that the entropy of adversarial patches is higher than that of clean images is not entirely consistent with reality. 
LGS provides a certain degree of defense against adversarial patches in YOLOv2 and YOLOv3, but it fails to defend against them in YOLOv4. 
As for OS, it naturally has the ability to resist the adversarial patch placed in the corner of the image since it defends patches based on sliding masks. It exhibits inferior defense performance against adversarial patches in YOLOv2 and YOLOv3 compared to NutNet, but it performs best in YOLOv4. This may be because YOLOv4 has a stronger detection capability, while YOLOv2 and YOLOv3 are more prone to generating redundant false positive detection boxes due to sliding window effects. Despite OS's higher effectiveness than NutNet in defending against the untargeted HA on YOLOv4, the adopted sliding mask and unionizing boxes are extremely time-consuming, making it unable to meet the real-time detection requirements of the object detector. The efficiency of defense methods will be discussed in Section~\ref{sec:efficiency}.

\vspace{3pt}
\noindent\textbf{Appearing Attack. }
Utilizing the method in~\cite{seeingisnotbelieving}, we generate two patches as the stop sign and the person based on the YOLOv2 detector and place them in 1,000 cropped images randomly sampled from the test set of the KITTI dataset to build the two datasets for evaluation. 
Note that although the patches are generated based on the YOLOv2 detector, the ``stop sign'' patch can successfully transfer to attack the other detectors while the ``person'' patch can succeed on the other detectors except Faster RCNN. Therefore, we evaluate all six detectors on the two datasets. The experimental results are shown in Table~\ref{tab:aa}. Since SSD is trained on the VOC dataset with no stop sign category, we only evaluate the ``person'' patch on SSD. Considering that the goal of the AA is to make the detector incorrectly detect the patch as a specific object, a lower AP indicates a better defensive performance after applying the defense.

NutNet has demonstrated excellent performance in defending against Appearing Attacks on all the testing models by decreasing the $\text{AP}_{0.5}$ to almost 0. After passing through NutNet, most of the pixels of the AA patch in the image are filtered out, making it difficult for the detector to detect the patch as a specific class, as shown in Figure~\ref{fig:digital example}. 
In contrast, SAC and OS exhibit minimal effectiveness in countering AA. Jedi and LGS, while showing some efficacy in reducing AP, demonstrate insufficient effectiveness. In most cases, the model maintains a relatively high detection accuracy for these AA patches. Only when defending against the ``stop sign'' patch for DETR do these two methods show a noticeable defensive effect, indicating the limited generality of Jedi and LGS.

\begin{table}[t]
    \centering
    \caption{Performance ($\text{AP}_{0.5}$) of NutNet against AdvPatch~\cite{Thys_2019_CVPR_Workshops} of different sizes. 0.2 means the patch height is 0.2 times the length of the diagonal of the corresponding bounding box of the object.}
    \begin{tabular}{c|c|c|c|c|c}
    \hline
        Size & 0.2 & 0.25 & 0.3 & 0.35 & 0.4 \\
        \hline
        AdvPatch w/o NutNet & 0.327 & 0.136 & 0.066 & 0.031 & 0.023 \\
        AdvPatch w/ NutNet & 0.762 & 0.721 & 0.656 & 0.498 & 0.344 \\
    \hline
        Gray Patch & 0.798 & 0.751 & 0.672 &  0.524 & 0.334 \\
    \hline
        Ground Truth & \multicolumn{5}{c}{0.847} \\
    \hline
    \end{tabular}
    \label{tab:ap sizes}
\end{table}

\vspace{3pt}
\noindent \textbf{Impact of different patch sizes.}
We extend our experiments to investigate the impact of various sizes of AdvPatch~\cite{Thys_2019_CVPR_Workshops} 
on the $\text{AP}_{50}$ of YOLOv2 detection on INRIA, as detailed in Table~\ref{tab:ap sizes}. Here, size 0.2 corresponds to the size utilized in Section~\ref{sec:defense digital}. \new{We also evaluate the detection accuracy of the object detector when adversarial patches are replaced with gray patches of the same size. 
And the ground truth represents the $\text{AP}_{50}$ on the clean dataset. }
While enlarging the patch size does enhance the attack success rate, the detection performance with NutNet consistently aligns with the performance on gray patches, exhibiting a marginal difference of only approximately 2-3\%. This finding suggests that NutNet can effectively detect patches of various sizes.

\subsection{Robustness}
\label{sec:robustness}

\begin{table}[]
    \centering
    \caption{Performance ($\text{AP}_{0.5}$) of adaptive attacks against NutNet. NN denotes NutNet. }
    \resizebox{\columnwidth}{!}{
    \begin{tabular}{c|p{0.9cm}p{0.8cm}|p{0.9cm}p{0.8cm}|p{0.9cm}p{0.8cm}}
    \hline
        \multirow{2}{*}{Model} & \multicolumn{2}{c|}{YOLOv2} & \multicolumn{2}{c|}{YOLOv3} & \multicolumn{2}{c}{YOLOv4} \\
        & w/o NN & w/ NN & w/o NN & w/ NN & w/o NN & w/ NN \\
        \hline
        Clean & \multicolumn{2}{c|}{0.847} & \multicolumn{2}{c|}{0.961} & \multicolumn{2}{c}{0.955} \\
        AdvPatch (a)~\cite{Thys_2019_CVPR_Workshops} & 0.310 & 0.780 & 0.206 & 0.837 & 0.417 & 0.829 \\
        Large $\alpha>1$ & 0.780 & 0.796 & 0.630 & 0.805 & 0.660 & 0.849 \\
        Small $\alpha<1$ & 0.520 & 0.786 & 0.429 & 0.769 & 0.456 & 0.852 \\
    \hline
    \end{tabular}}
    \label{tab:adaptive attack}
\end{table}


\vspace{3pt}
\noindent\textbf{Adaptive Attack. }
In this section, we evaluate whether NutNet can successfully defend against adaptive attacks. We utilize an autoencoder to filter out non-clean images, \emph{i.e.} adversarial patches. Given that the autoencoder is a convolutional neural network, it is also susceptible to adversarial attacks. In our threat model, an attacker, aware of our defense setup, might execute adaptive attacks concurrently targeting both NutNet and the object detector. The ability to withstand adaptive attacks is a crucial aspect in assessing the practicality of a defense mechanism.

We assume that the attacker can obtain all the information about the autoencoder we use. Therefore, when generating adversarial patches, the attacker can add another loss term targeting NutNet, whose goal is to minimize the reconstruction loss after the adversarial patch passes through the autoencoder. 
The objective function for the adaptive attack is depicted as follows,

\begin{equation}
\label{equ:adaptive attack}
    \min L(D(A(p,x)), y_{gt}) + \alpha Dist(\mathcal{E}(A(p,x)), A(p,x))
\end{equation}

\noindent where $p$ is the adversarial patch, $(x,y_{gt})$ represents the clean image and its ground-truth annotations, $D$ is the victim object detector, $L$ is the objective function of adversarial attack and $A$ is the patch apply function which applies the patch to the clean image $x$. Note that the attacker cannot optimize a patch directly by minimizing $L$ of the reconstructed image $\mathcal{E}(A(p,x))$, since $\mathcal{E}$ consists of reconstruction and dual-mask generation, where the latter is non-differentiable.

We utilize the aforementioned formulation to generate adaptive adversarial patches for hiding persons against the YOLOv2-v4 detectors. Subsequently, testing is conducted on the INRIA dataset, following the configurations outlined in Section~\ref{sec:defense digital}. Under the condition of a larger $\alpha$, the adaptive patches indeed exhibit a higher evasion rate against NutNet. However, compared to the adversarial patches employed in the previous experiments, the attack performance of these adaptive patches is notably inferior. Conversely, when a smaller $\alpha$ is set, the adaptive patches still maintain a reasonable level of attack effectiveness but remain unable to bypass NutNet. The experimental results are presented in Table~\ref{tab:adaptive attack}.

Without any defense, the vanilla YOLOv2-v4 detectors exhibit $\text{AP}_{0.5}$ of only 0.31, 0.206, and 0.417 respectively when confronted with the original adversarial patches. However, when faced with the adaptive patches of a relatively large $\alpha$ (over 1.0), they achieve $\text{AP}_{0.5}$ of 0.780, 0.630, and 0.660, much higher than the original patches. This indicates that these adaptive attack patches, designed to evade NutNet, have sacrificed a significant portion of their adversarial effectiveness. Moreover, it remains challenging for these adaptive patches to completely bypass NutNet, and there might still be a small fraction filtered out. Therefore, under the defense of NutNet, the detection accuracy can still experience a slight improvement, reaching an $\text{AP}_{0.5}$ of 0.8 or higher.
When $\alpha$ is set smaller, the adaptive patches maintain a certain level of attack effectiveness, causing vanilla YOLOv2-v4 detectors to achieve $\text{AP}_{0.5}$ of only 0.520, 0.429, and 0.456. Nevertheless, they still cannot bypass NutNet. In the presence of NutNet defense, all three models attain $\text{AP}_{0.5}$ of 0.786, 0.769, and 0.852 respectively.

\begin{figure}[]
    \centering
    \includegraphics[width=0.9\linewidth]{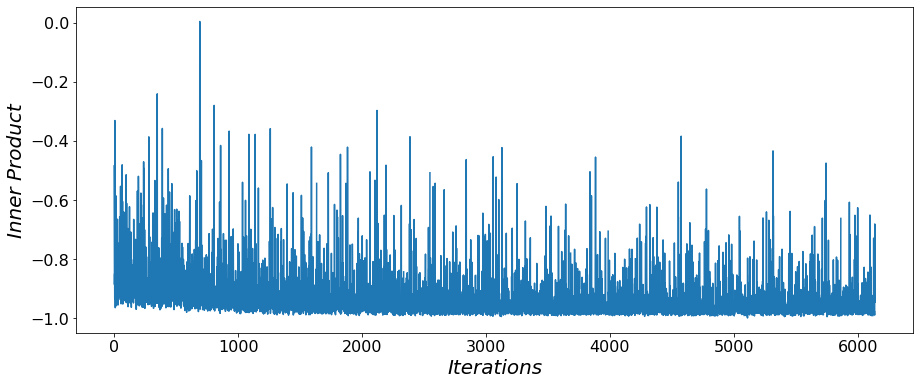}
    \caption{The inner product between the gradient of attacking the object detector and the gradient of bypassing NutNet.}
    \label{fig:grad inner product}
    \vspace{-0.3cm}
\end{figure}

We analyze why adaptive patches cannot simultaneously attack the object detector and bypass NutNet. Using Equation~\ref{equ:adaptive attack}, we generate an adaptive adversarial patch, recording adversarial attack loss against the object detector and reconstruction loss against NutNet iteratively. We then calculate and normalize the gradients of the adversarial patch for these two losses, computing their inner product, as shown in Figure~\ref{fig:grad inner product}. In general, we observe that the inner product is consistently negative during the iterative process, indicating an angle consistently greater than $90^\circ$ between the two gradients. This implies that the adaptive patch, aiming to reduce the adversarial attack loss, unavoidably increases the reconstruction loss against NutNet, and vice versa. Consequently, it is challenging for this adversarial patch to simultaneously attack the object detector and bypass NutNet.

SAC, which also utilizes an extra model for defense, undergoes adaptive attacks for comparative evaluation. Employing objectives akin to Equation~\ref{equ:adaptive attack}, we optimize adversarial patches customized for YOLOv3. The results reveal a further reduction in detection accuracy under SAC defense, dropping to 0.437, while ours is 0.805/0.769. 
This indicates that the resilience of methods relying on conventional segmentation models for adversarial patch defense proves insufficient against adaptive attacks.

\begin{figure}[t]
    \centering
    \begin{subfigure}{0.85\linewidth}
        \centering
        \includegraphics[width=\linewidth]{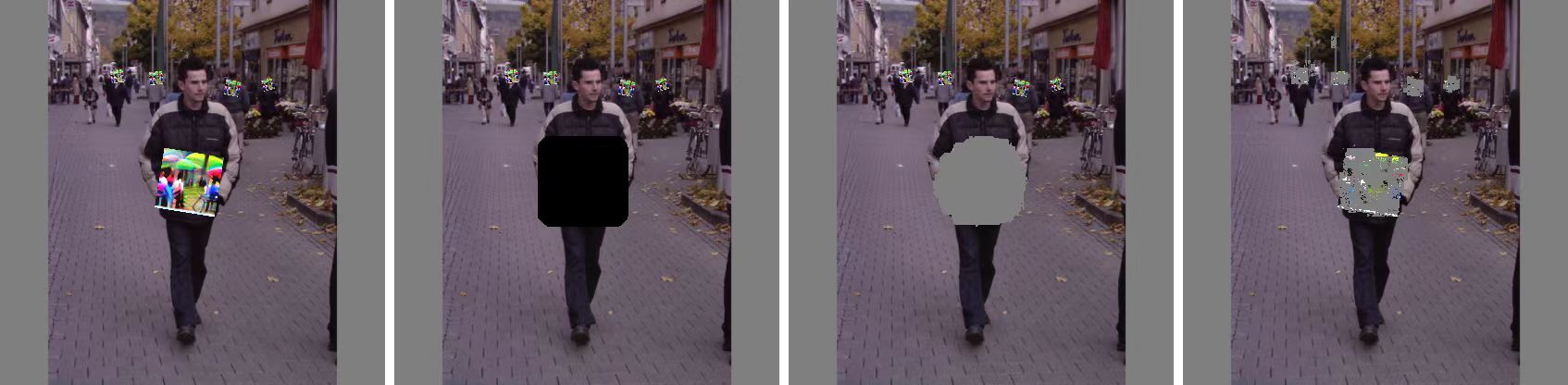}
    \end{subfigure}
    \vspace{2pt}\vfill
    \begin{subfigure}{0.85\linewidth}
        \centering
        \includegraphics[width=\linewidth]{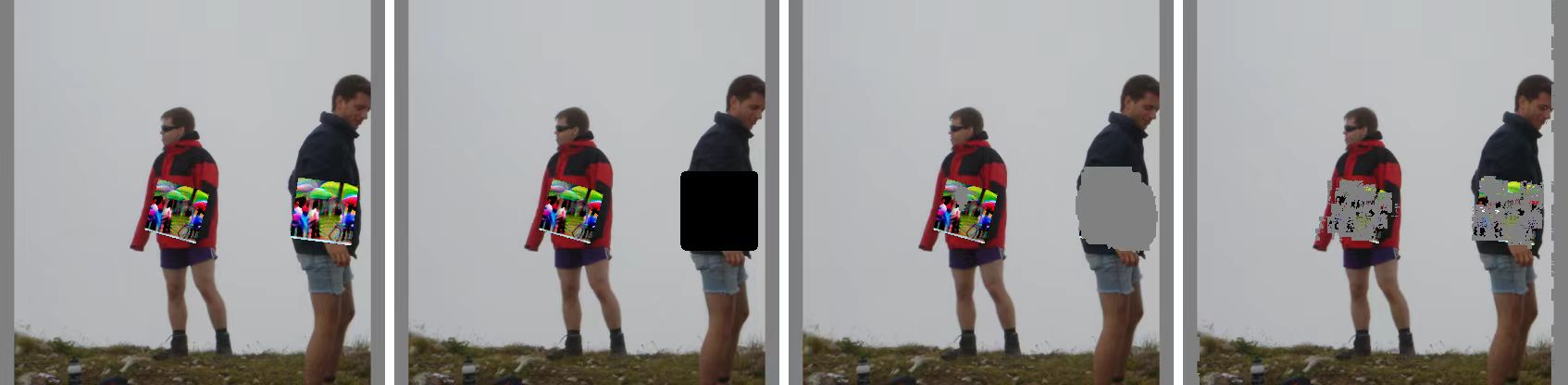}
    \end{subfigure}
    \vspace{2pt}\vfill
    \begin{subfigure}{0.85\linewidth}
        \centering
        \includegraphics[width=\linewidth]{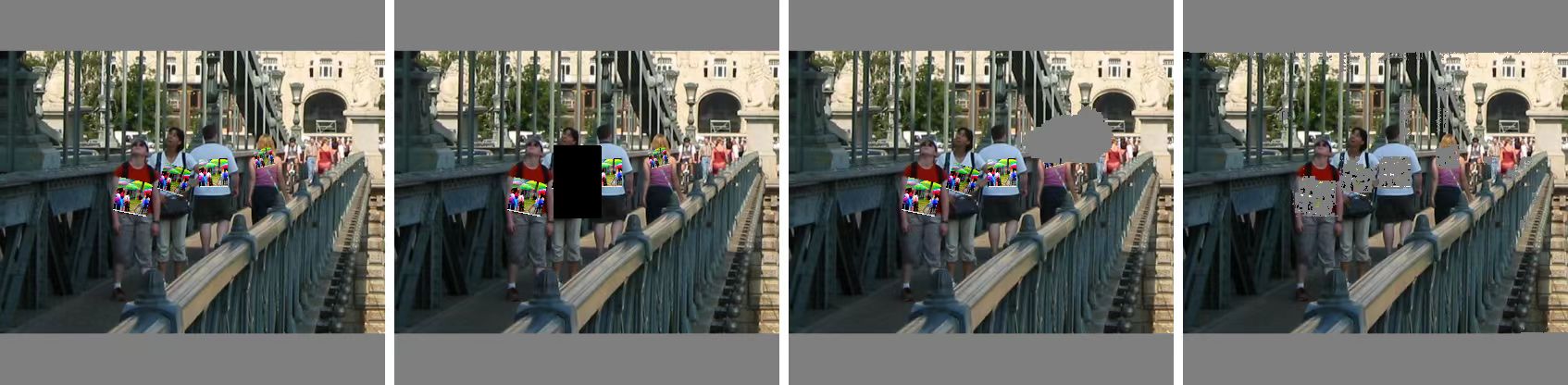}
    \end{subfigure}
    \vspace{1pt}\vfill
    \caption{Examples of the defensive performance of SAC (column 2), Jedi (column 3) and NutNet (column 4). The first column shows the original patched images. Note that SAC uses black masks while Jedi and NutNet use gray ones.}
    \label{fig:mask examples}
    \vspace{-0.3cm}
\end{figure}

\vspace{3pt}
\noindent\textbf{Patch with Transformations. }
Evaluating the effectiveness of adversarial defenses against patches with various transformations (\emph{e.g.}, rotation, scaling) is crucial, given attackers' use of irregular, discrete, and small-sized adversarial patches~\cite{zhu2021easily} to bypass patch detection methods.
We create a dataset with diverse transformations using AdvPatch~\cite{Thys_2019_CVPR_Workshops} and the INRIA training set. We assess the detection and masking performance of NutNet on this dataset. Since SAC~\cite{Liu_2022_CVPR} and Jedi~\cite{Tarchoun_2023_CVPR} also defend against patches using masking, we conduct comparative tests. We measure the overlap ratio between the generated masks for each image and the ground truth masks as follows: $\lVert m_d \odot m_{gt} \rVert / \lVert m_{gt} \rVert$, where $m_d$ and $m_{gt}$ represent the generated masks and ground truth masks respectively, and $\odot$ is the element-wise production. SAC and Jedi achieve 0.53 and 0.39 respectively. As a comparison, NutNet achieves an overlap ratio of 0.97, which is much better than SAC and Jedi.

Figure~\ref{fig:mask examples} indicates some examples, which display original images in the first column, SAC and Jedi performance in the second and third columns, and NutNet performance in the fourth column. We can observe that NutNet can accurately capture adversarial patches with various transformations, even when these patches are reduced to almost invisible states (see the first row of Figure ~\ref{fig:mask examples}). In contrast, SAC exhibits detection failures for small adversarial patches in the image, and the output masks often do not align well with the shape of the adversarial patches. Jedi's detection performance is not entirely stable, with output mask shapes appearing irregular and, at times, failing to accurately locate the adversarial patches (refer to the third row of Figure~\ref{fig:mask examples}). More examples can be found in Appendix~\ref{sec:appendix more digital}.

We further analyze the factors contributing to the differences in the effectiveness of these defense methods. Jedi detects adversarial patches based on entropy levels, assuming higher entropy in adversarial patches than in clean images. However, we find that high entropy does not necessarily indicate adversarial patches. Additionally, operations like scaling or blurring can further reduce patch entropy. Relying solely on entropy levels to identify adversarial patches in a specific area is challenging. 
SAC's limited defensive performance may stem from two aspects. On one hand, during training, SAC might have encountered larger-sized adversarial patches, often with rectangular shapes, leading to additional information incorporation (\emph{i.e.}, overfitting). This aligns with our emphasis in Section~\ref{sec:back_defense} on the consequences of strong coupling between the defense model and the encountered training models. On the other hand, SAC conducts detection across the entire image, and small patches may only constitute a small fraction, posing a challenge for detection. Images passing through convolutional layers can result in smaller adversarial patches and their surrounding clean image regions being sampled into the same feature map area, diluting adversarial patch features and making it challenging to distinguish them from clean images.

\begin{figure}[t]
    \centering
    \includegraphics[width=0.9\linewidth]{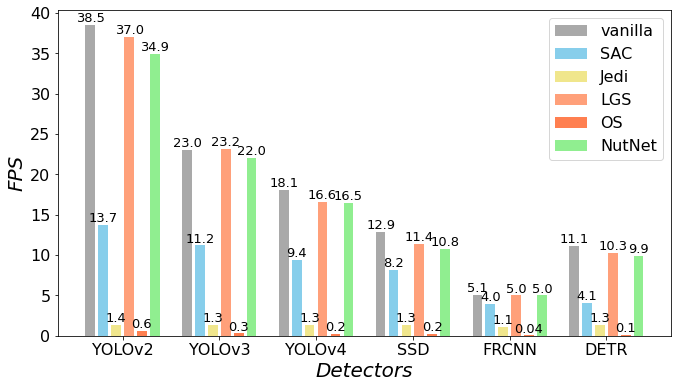}
    \caption{Efficiency (FPS) of different defense methods for different detectors running on KITTI dataset.}
    \label{fig:efficiency}
\end{figure}

\subsection{Efficiency}
\label{sec:efficiency}
In this section, we evaluate prediction efficiency using Frames Per Second (FPS) for six models under different defense methods. Given the frequent use of object detection models in autonomous vehicles, we utilize the KITTI dataset~\cite{doi:10.1177/0278364913491297}, a well-known dataset for evaluating vision algorithms in autonomous driving scenarios. We randomly select 1000 images from the KITTI test set and input them into various object detectors, recording their detection FPS with different defense mechanisms, as illustrated in Figure~\ref{fig:efficiency}.

Experimental results indicate that LGS and NutNet have little impact on the inference efficiency of the vanilla model, with only 5.5\% and 8.6\% overhead respectively. Both LGS and NutNet involve adding another small module for image pre-processing before the detection model, which has a very simple architecture and a very short processing time. Such a small additional overhead hardly affects the real-time detection capability of the detector. 
The U-Net used in SAC has a large number of parameters (over 1M), leading to a longer forward inference time. As a result, the detection speed of the model under SAC defense is typically reduced by around 50\%. Jedi requires the image to be divided into blocks of different sizes, and it calculates the two-dimensional entropy for generating masks. Additionally, Jedi uses an autoencoder to cluster discrete masks, resulting in a significant computational overhead. 
For OS, however, its sliding mask and unionizing boxes operations are highly time-consuming, resulting in significant processing time for detecting a single image, often taking several seconds. 
In conclusion, LGS and NutNet are capable of meeting the real-time requirements of defending object detectors, while the others are too time-consuming and cannot fulfill real-time demands. 

\subsection{Comparison with Existing Defenses}
\label{sec:compare}
The experiments conducted in Sections~\ref{sec:defense digital}, \ref{sec:robustness} and \ref{sec:efficiency} demonstrate that NutNet surpasses existing methods in terms of both effectiveness and efficiency. NutNet exhibits robust defense capabilities against both HA and AA across all tested models, with minimal additional overhead. A summative comparison is shown in Table~\ref{tab:comparison summary}.

SAC provides a slight enhancement in detection accuracy for clean images (Table~\ref{tab:uha}). However, its effectiveness against adversarial patches is limited, and the defensive outcome varies (Table~\ref{tab:tha} and~\ref{tab:aa}). 
Jedi demonstrates slight defensive capabilities against AA (Table~\ref{tab:aa}), but its defensive effectiveness is inconsistent for HA (Table~\ref{tab:tha} and~\ref{tab:uha}). Sometimes it provides a defense, while sometimes it may have a negative impact, challenging Jedi's assumption that adversarial patch entropy surpasses that of clean images.
OS defends against adversarial patches in corners using sliding masks but shows mixed results. It significantly impacts YOLOv2, SSD, and Faster RCNN but performs better in YOLOv4 (Table~\ref{tab:tha}). The method's high computational cost, primarily due to sliding masks, raises concerns about real-time detection (Figure~\ref{fig:efficiency}). 
LGS has a certain level of defense against various types of attacks (Table~\ref{tab:tha} and~\ref{tab:uha}). However, its effectiveness in defending against AA is unstable, sometimes exhibiting defense capabilities and at other times further enhancing the attack performance (Table~\ref{tab:aa}).

From the perspective of efficiency in Figure~\ref{fig:efficiency}, NutNet and LGS introduce an additional overhead of 5.5\% and 8.6\% respectively, which can be considered negligible and will not impact the real-time performance of the detector. However, other methods fail to ensure real-time performance of the object detector. SAC results in a detection speed reduction of approximately 50\%, while Jedi typically only maintains a detection speed of around 1 FPS. OS reduces the efficiency to around 1\% and typically takes several seconds to complete detection for a single image, making it unsuitable for real-time use.

\begin{table}[]
    \centering
    \caption{Comparative analysis of generalization, robustness, and efficiency. \ding{73}\ding{73} indicates generally limited performance, \ding{72}\ding{73} suggests occasional effectiveness, and \ding{72}\ding{72} denotes consistently good performance.}
    \resizebox{\columnwidth}{!}{
    \begin{tabular}{c|c|c|c|c|c}
    \hline
        Defense & SAC~\cite{Liu_2022_CVPR} & Jedi~\cite{Tarchoun_2023_CVPR} & LGS~\cite{DBLP:journals/corr/abs-1807-01216} & OS~\cite{xiang2022objectseeker} & NutNet \\
    \hline
        Generalization to targeted HA & \ding{72}\ding{73} & \ding{72}\ding{73} & \ding{72}\ding{73} & \ding{73}\ding{73} & \ding{72}\ding{72} \\
        Generalization to untargeted HA & \ding{72}\ding{72} & \ding{72}\ding{73} & \ding{72}\ding{73} & \ding{72}\ding{73} & \ding{72}\ding{72} \\
        Generalization to AA & \ding{73}\ding{73} & \ding{72}\ding{73} & \ding{72}\ding{73} & \ding{73}\ding{73} & \ding{72}\ding{72} \\
        Robustness & \ding{72}\ding{73} & \ding{73}\ding{73} & - & - & \ding{72}\ding{72} \\
        Efficiency & \ding{72}\ding{73} & \ding{73}\ding{73} & \ding{72}\ding{72} & \ding{73}\ding{73} & \ding{72}\ding{72} \\
    \hline
    \end{tabular}}
    \label{tab:comparison summary}
    \vspace{-0.3cm}
\end{table}

\begin{table}[]
 \caption{Success rates of the physical attack against YOLOv2.}
    \centering
    \resizebox{0.9\columnwidth}{!}{
    \begin{tabular}{cc|cc}
        \hline
         \multicolumn{2}{c|}{\multirow{2}{*}{Attack}} &  \multicolumn{2}{c}{Success rate} \\
         & & w/o NutNet & w/ NutNet \\
        \hline
         \multirow{2}{*}{Targeted HA} & \multicolumn{1}{|c|}{AdvPatch (a)~\cite{Thys_2019_CVPR_Workshops}} & \multicolumn{1}{c|}{83.0\%} & 0.7\%\\
         & \multicolumn{1}{|c|}{NaturalPatch \cite{Hu_2021_ICCV}} & \multicolumn{1}{c|}{39.0\%} & 1.0\%\\
        \hline
         \multirow{1}{*}{Untargeted HA} & \multicolumn{1}{|c|}{PAPatch~\cite{lee2019physical}} & \multicolumn{1}{c|}{74.9\%} & 0.3\%\\
        \hline
         \multirow{2}{*}{AA} & \multicolumn{1}{|c|}{stop sign} & \multicolumn{1}{c|}{96.3\%} & 0\%\\
         & \multicolumn{1}{|c|}{person} & \multicolumn{1}{c|}{98.7\%} & 5.6\%\\
        \hline
    \end{tabular}}
    \label{tab:physical}
    \vspace{-0.3cm}
\end{table}

\subsection{Effectiveness in the Physical World}
\label{sec:defense physical}
We also conduct tests on the effectiveness of defense against HA and AA in the physical world. Demos can be found at: \url{https://sites.google.com/view/nutnet}.
The goal of untargeted HA is to hide all objects in the scene. However, measuring the success rate of the attack based on this criterion is not ideal because not all objects can be consistently hidden. Therefore, we measure the attack success rate by evaluating whether it can hide the person in the scene, similar to targeted HA. 
Specifically, we print out different types of adversarial patches and carry out physical attacks. We record videos of the camera and the object performing relative motion and input them into the detector. Each patch's video lasted for more than 40 seconds (30 FPS). We calculate the success rate of the adversarial attacks in each video, where the success rate is defined as the percentage of frames in which the attack succeeded out of the total frames. Then, we apply NutNet to process the videos before inputting them into the detector, and we calculate the attack success rate in each video again. 
Each patch is evaluated on the model it is specifically targeted against, rather than employing transferability attacks. 
The experimental results are shown in Table~\ref{tab:physical}. 
Regardless of the type of attack, NutNet is able to significantly reduce the attack success rate to a very low level. Note that the original attack success rate of NaturalPatch\cite{Hu_2021_ICCV} is already low, which is consistent with the evaluation in the digital world in Section \ref{sec:defense digital}. More physical evaluations will be demonstrated in Appendix~\ref{sec:appendix more physical}.

\begin{table}[]
\caption{Performance ($\text{AP}_{0.5}$) of NutNet with different block sizes under different attacks against YOLOv3 detector. NutNet-13, NutNet-26 and NutNet-52 represent the block sizes 13, 26 and 52 used by NutNet.}
    \centering
    \resizebox{\columnwidth}{!}{
    \begin{tabular}{c|c|c|c|c|c}
        \hline
        \multicolumn{2}{c|}{Attack} & Vanilla & NutNet-13 & NutNet-26 & NutNet-52 \\
        \hline
        \multirow{2}{*}{Targeted HA} & clean & 0.961 & 0.951 & 0.961 & 0.940 \\
         & ~\cite{Thys_2019_CVPR_Workshops}$^*$ & 0.206 & 0.837 & 0.835 & 0.736\\
        \hline
         \multirow{2}{*}{Untargeted HA} & clean & 0.513 & 0.495 & 0.511 & 0.513 \\
         & ~\cite{lee2019physical} & 0.204 & 0.425 & 0.438 & 0.431 \\
        \hline
        \multirow{2}{*}{AA} & stop sign & 0.806 & 0.081 & 0.007 & 0.058 \\
         & person & 0.952 & 0.002 & 0.031 & 0.002 \\
        \hline
    \end{tabular}}
\label{tab:abla_size}
\vspace{-0.3cm}
\end{table}

\subsection{Ablation Study}
\label{sec:abla}
In this section, we conduct tests of different configurations on YOLOv2 and YOLOv3 to defend against the HA and AA patches to evaluate the defense performance of different configurations. 

\vspace{3pt}
\noindent\textbf{Impact of the Block Size. }
We apply NutNet of different block sizes including $13\times13$, $26\times26$ and $52\times52$ to YOLOv3 and evaluate the defense performance. Experimental results are shown in Table~\ref{tab:abla_size}. NutNet-13, NutNet-26, and NutNet-52 are all effective in defending against AA.
Due to the larger size of patches of the untargeted HA, NutNet with larger block sizes can effectively filter out the patches with low false positives. Therefore, NutNet-26 and NutNet-52 perform better than NutNet-13 in terms of effectively filtering the patches. In targeted HA, the patches are relatively smaller, making it less likely for NutNet with large block sizes to filter out small-sized patches. Therefore, NutNet-13 and NutNet-26 perform better than NutNet-52 in targeted HA scenarios. 

\vspace{3pt}
\noindent\textbf{Impact of the Mask Type. }
As mentioned in Section \ref{sec:defending approach}, NutNet uses a coarse-grained mask $m_1$ and a fine-grained mask $m_2$ in combination to process the image. Here, we evaluate the defensive effect of using only one mask. Experimental results are shown in Table~\ref{tab:abla_mask}. Using only one of the masks results in a much lower defensive effect compared to using them in combination. When using only the coarse-grained mask $m_1$, it is possible to mask the entire block containing only a small patch region, thereby affecting the model's detection accuracy. On the other hand, using only the fine-grained mask $m_2$ may result in the image being filled with discrete small masked pixels, which also affects the model's detection accuracy.

\begin{table}[]
\caption{Performance ($\text{AP}_{0.5}$) of NutNet with different masks under different attacks against YOLOv3 detector. $m_1\odot m_2$, $m_1$ and $m_2$ represent using different masks.}
    \centering
    \resizebox{\columnwidth}{!}{
    \begin{tabular}{c|c|c|c|c|c}
        \hline
        \multicolumn{2}{c|}{Attack} & Vanilla & $m_1\odot m_2$ & $m_1$ & $m_2$\\
        \hline
        \multirow{2}{*}{Targeted HA} & clean & 0.961 & 0.951 & 0.912 & 0.895 \\
         & ~\cite{Thys_2019_CVPR_Workshops}$^*$ & 0.206 & 0.837 & 0.763 & 0.687\\
        \hline
         \multirow{2}{*}{Untargeted HA} & clean & 0.513 & 0.495 & 0.432 & 0.314 \\
         & ~\cite{lee2019physical} & 0.204 & 0.425 & 0.416 & 0.285 \\
        \hline
        \multirow{2}{*}{AA} & stop sign & 0.806 & 0.081 & 0.000 & 0.123 \\
         & person & 0.952 & 0.002 & 0.000 & 0.001 \\
        \hline
    \end{tabular}}
\label{tab:abla_mask}
\vspace{-0.3cm}
\end{table}

\begin{table}[]
\caption{\new{Performance ($\text{AP}_{0.5}$) of NutNet with different thresholds $\kappa$ under no attacks, HA (AdvPatch~\cite{Thys_2019_CVPR_Workshops}) and AA (the person patch) against YOLOv2 detector. }}
\centering
    \resizebox{\columnwidth}{!}{
    \begin{tabular}{c|ccc|ccc|ccc}
    \hline
    \multirow{2}{*}{\diagbox[]{$\kappa_1$}{$\kappa_2$}} & \multicolumn{3}{c|}{0.1} & \multicolumn{3}{c|}{0.2} & \multicolumn{3}{c}{0.3} \\
    & clean & HA & AA & clean & HA & AA & clean & HA & AA \\
    \hline
    0.04 & 0.800 & 0.648 & 0.0 & 0.858 & 0.724 & 0.0 & 0.868 & 0.763 & 0.039 \\
    0.08 & 0.835 & 0.708 & 0.0 & 0.858 & 0.764 & 0.002 & 0.866 & 0.784 & 0.121 \\
    0.12 & 0.855 & 0.734 & 0.002 & 0.868 & 0.786 & 0.014 & 0.873 & 0.795 & 0.212 \\
    0.16 & 0.865 & 0.732 & 0.002 & 0.877 & 0.783 & 0.061 & 0.876 & 0.786 & 0.330 \\
    0.20 & 0.876 & 0.761 & 0.057 & 0.875 & 0.782 & 0.199 & 0.877 & 0.794 & 0.457 \\
    \hline
    \end{tabular}}
\label{tab:abla_k1k2}
\vspace{-0.3cm}
\end{table}

\vspace{3pt}
\noindent \new{\textbf{Impact of the Threshold $\kappa$. }
The threshold $\kappa$ affects NutNet's sensitivity to suspicious blocks. Therefore, we also tested the detection accuracy of YOLOv2 with different NutNet thresholds, as shown in Table~\ref{tab:abla_k1k2}. 
Overall, as $\kappa_1$ and $\kappa_2$ increase, the detection accuracy on both clean images and images with HA patches improves, likely due to the larger thresholds resulting in a very low false positive rate for NutNet. However, larger $\kappa_1$ and $\kappa_2$ lead to poorer defense against AA patches, as the masking rate for these patches decreases, causing more false negatives.
In summary, selecting appropriate thresholds is crucial for balancing the false positive of defending against HA and false negative of defending against AA effectively.}

\section{Discussion}
\label{sec:discussion}

\noindent \new{\textbf{Impact of Noise Distribution. } 
In Section~\ref{sec:training approach}, we simulate non-clean distribution images by adding randomly sampled Gaussian noise blocks to normal images. One might question whether using Gaussian noise could cause NutNet only to detect patches from Gaussian distributions, potentially allowing patches from non-Gaussian distributions to evade detection. However, this does not affect NutNet's detection capabilities. Patches don't have to follow the same distribution as the added noises. Most patches in our experiments don't follow Gaussian distributions, yet NutNet still works. Adding noise prevents the autoencoder's reconstruction capability from transferring to non-clean distributions, thereby enhancing its ability to filter out-of-distribution patterns. }

\new{We also evaluate the defensive performance trained with noise blocks sampled from a uniform distribution. Experimental results indicate that NutNet trained with uniform distribution noise achieved comparable effectiveness to that trained with Gaussian noise. This suggests that NutNet's defensive capabilities are not tied to a specific distribution of noise blocks.}

\vspace{3pt}
\noindent \new{\textbf{Impact of Attacker's Configurations. }
In Section~\ref{sec:threat model}, we introduced the attacker's capabilities in the threat model, stating that the attacker can manipulate the attack type, size, shape, position, and number of patches. Here, we will summarize how these attack configurations affect the defense performance of NutNet. 
Experimental results in Section~\ref{sec:defense digital} and Section~\ref{sec:defense physical} demonstrate that NutNet can defend against different types of attacks, including adversarial patches of various sizes which are used to simulate different distances, positions (target HA patches overlapping with the target, non-target HA patches in the corners of the image, and AA patches in arbitrary positions within the image), and the number (the number of target HA patches depends on the number of target objects). Additionally, Figure~\ref{fig:extractor} 
in the Appendix shows that NutNet can detect adversarial patches of different shapes, including rectangular, circular, and other shapes formed by perspective transformations. Overall, despite the attacker's ability to manipulate various attack configurations, NutNet can successfully defend against them.}

\begin{table}[t]
\caption{\new{Performance ($\text{AP}_{0.5}$) of YOLOv2 detector on INRIA dataset under AdvPatch of different masking rates.}}
    \centering
    \resizebox{0.9\columnwidth}{!}{
    \begin{tabular}{c|cccccc}
        \hline
        \multirow{2}{*}{patch} & \multicolumn{6}{c}{mask rate} \\
        & 0 & 0.5 & 0.6 & 0.7 & 0.8 & 0.9 \\
        \hline
        AdvPatch (a) & 0.218 & 0.722 & 0.759 & 0.773 & 0.783 & 0.783 \\ 
        \hline
    \end{tabular}}
\label{tab:random mask}
\vspace{-0.3cm}
\end{table}

\vspace{3pt}
\noindent \new{\textbf{Impact of False Negative. }
In Section~\ref{sec:defending approach} we introduce DualMask to reduce NutNet's false positives, which might raise concerns about NutNet's false negatives. Therefore, we also evaluate the attack performance when adversarial patches are only partially masked. Specifically, we used the same setup in Section~\ref{sec:defense digital}, applying varying degrees of random masking to AdvPatch and testing YOLOv2's detection accuracy on the INRIA dataset. The experimental results are shown in Table~\ref{tab:random mask}. Even with only 50\% masking of the patch, the attack performance of the adversarial patch drops significantly. As mentioned in Section~\ref{sec:robustness} NutNet's mask rate for AdvPatch can reach 97\%. Therefore, even if some adversarial patch pixels are missed after NutNet's defense (\textit{i.e.}, false negatives), they are almost incapable of successfully attacking.}

\vspace{3pt}
\noindent \textbf{Towards Better Precision. }
Since the adversarial patch is directly placed in the image, it will inevitably obscure some of the original contextual information. 
Defending against adversarial patches in the image by filling them with gray color cannot make the detector output the same detection results as the clean image due to the loss of contextual information.
As demonstrated in Table~\ref{tab:tha} and~\ref{tab:uha}, NutNet can improve the average precision of the object detector on the datasets with patches, but it still falls short of the precision on the clean datasets.

\begin{table}[t]
\caption{Performance and efficiency of NutNet-13 with and without image inpainting under different patches against YOLOv2 detector on INRIA.}
    \centering
    \resizebox{0.85\columnwidth}{!}{
    \begin{tabular}{c|cc|cc|cc}
        \hline
        \multirow{2}{*}{patch} & \multicolumn{2}{c|}{vanilla} & \multicolumn{2}{c|}{w/o inpainting} & \multicolumn{2}{c}{w/ inpainting} \\
        & $\text{AP}_{50}$ & FPS & $\text{AP}_{50}$ & FPS & $\text{AP}_{50}$ & FPS \\
        \hline
        clean & 0.847 & 28.5 & 0.868 & 31.3 & 0.876 & 16.0 \\ 
        \cite{Thys_2019_CVPR_Workshops} & 0.310 & 36.8 & 0.759 & 41.0 & 0.803 & 17.7 \\ 
        \cite{DBLP:journals/corr/abs-1910-11099} & 0.497 & 35.4 & 0.754 & 40.7 & 0.810 & 17.6 \\ 
        \cite{10.1007/978-3-030-58548-8_1} & 0.431 & 36.1 & 0.760 & 40.9 & 0.808 & 17.5 \\ 
        \cite{Hu_2021_ICCV} & 0.637 & 34.0 & 0.708 & 40.3 & 0.729 & 17.3 \\ 
        \hline
    \end{tabular}}
\label{tab:inpainting}
\vspace{-0.3cm}
\end{table}

To compensate for the loss of contextual information caused by the mask, we can employ a generative model for recovering the masked regions (\emph{i.e.}, image inpainting). 
Inspired by~\cite{2021arXiv210805075C}, we try the PICnet~\cite{zheng2019pluralistic} (Pluralistic Image Completion) for inpainting the masked images and evaluate its improvement in defending against targeted HA on YOLOv2, as shown in Table~\ref{tab:inpainting}. The experimental results indicate that image inpainting does not provide much additional $\text{AP}_{50}$ improvement on clean datasets. However, on patched datasets, image inpainting can lead to an average $\text{AP}_{50}$ improvement of 4.25\%. Nevertheless, the additional overhead introduced by image inpainting is substantial. The average FPS for detection on different patched datasets with inpainting is only 43\% of the average FPS without inpainting.

Given limited hardware resources, there is a trade-off between precision and efficiency. Therefore, image inpainting is optional, depending on the requirement prioritization of detection precision and efficiency. 
If real-time performance is prioritized over detection precision, we can use NutNet only. Conversely, if detection precision is of higher importance, we can incorporate it into the defense pipeline.

\section{Conclusion}

In this paper, we presented a novel adversarial defense called \textit{NutNet} to protect the object detector from being attacked by adversarial patches. NutNet, functioning as a reconstruction-based autoencoder, is trained to differentiate between the distribution of clean images and that of non-clean images. As a result, NutNet can be used to detect adversarial patches, laying the foundation for subsequent patch masking.
Through experimental evaluation, NutNet demonstrates superior defense performance across various models, datasets, and attack types including HA and AA, with only 8\% overhead in the inference time of the detection system in most cases. Compared with baseline defenses, NutNet exhibits an average defense performance that is over 2.4 times and 4.7 times higher than existing approaches for HA and AA, respectively.

\section*{acknowledgements}
We thank all the anonymous reviewers for their constructive feedback.
The IIE authors are supported in part by National Natural Science Foundation of China (Grant No.92270204, 62302497 and 62302498) and Youth Innovation Promotion Association CAS.

\bibliographystyle{ACM-Reference-Format}
\bibliography{sample-base}


\appendix

\section{Details of DualMask Generation}
\label{sec:appendix dualmask}
\begin{figure}[t]
    \centering
    \begin{subfigure}{0.98\linewidth}
        \centering
        \includegraphics[width=\linewidth]{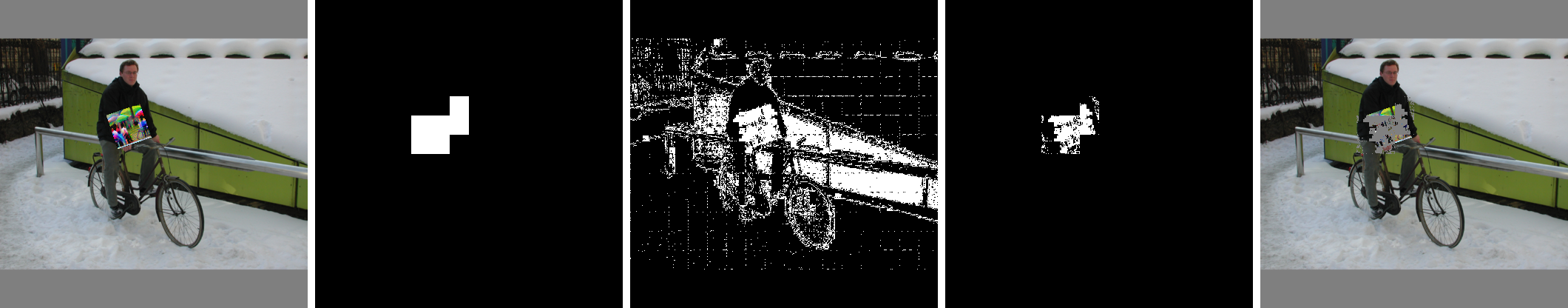}
    \end{subfigure}
    \vspace{2pt}\vfill
    \begin{subfigure}{0.98\linewidth}
        \centering
        \includegraphics[width=\linewidth]{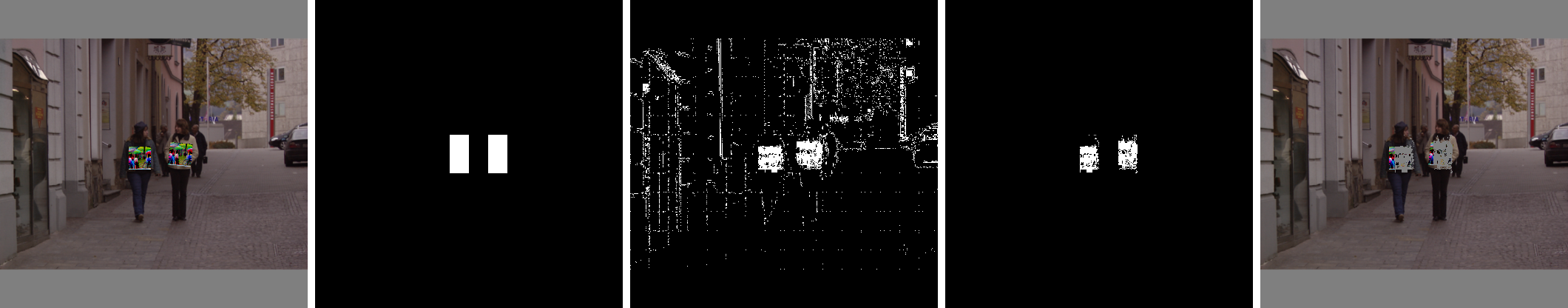}
    \end{subfigure}
    \vspace{2pt}\vfill
    \begin{subfigure}{0.98\linewidth}
        \centering
        \includegraphics[width=\linewidth]{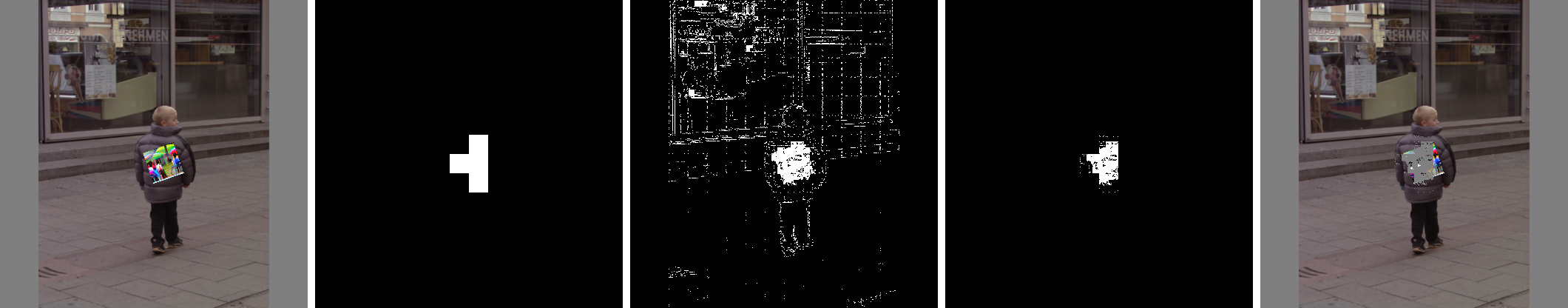}
    \end{subfigure}
    \vspace{2pt}\vfill
    \begin{subfigure}{0.98\linewidth}
        \centering
        \includegraphics[width=\linewidth]{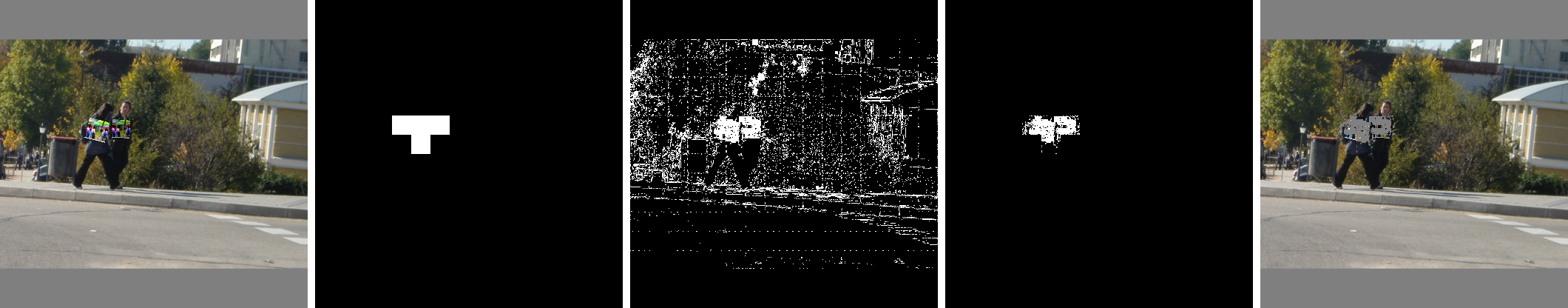}
    \end{subfigure}
    \vspace{2pt}\vfill
    \begin{subfigure}{0.98\linewidth}
        \centering
        \includegraphics[width=\linewidth]{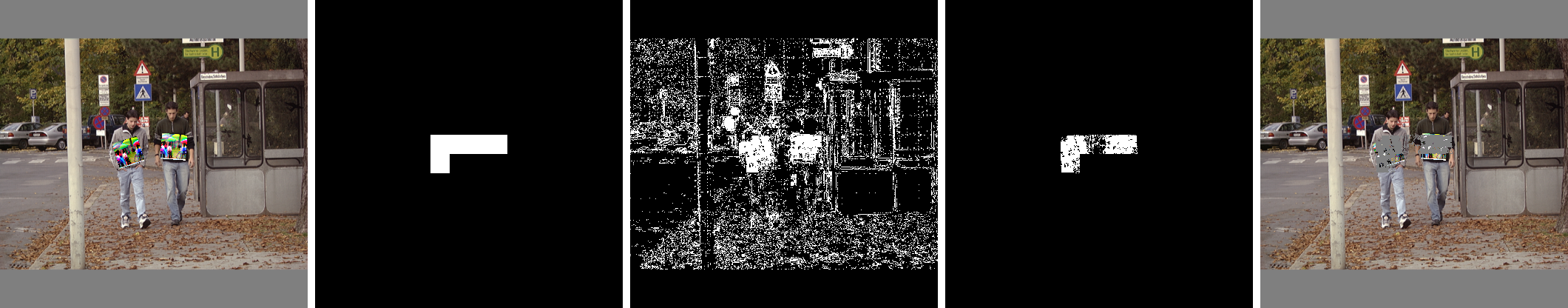}
    \end{subfigure}
    \vspace{1pt}\vfill
    \caption{Examples of DualMask Generation and its application. The sequence from left to right includes the original input (column 1), coarse-grained mask (column 2), fine-grained mask (column 3), the final mask (column 4), and the resulting output masked image (column 5).}
    \label{fig:dualmask example}
    \vspace{-0.4cm}
\end{figure}

In this section, we present specific instances of DualMask Generation, illustrated in Figure~\ref{fig:dualmask example}. Each row, from left to right, showcases the original input, the coarse-grained mask $m_1$, the fine-grained mask $m_2$, the final mask $m = m_1 \odot m_2$, and the resulting masked image. The coarse-grained mask $m_1$ is created based on the reconstruction error of blocks, resulting in a combination of square blocks. The fine-grained mask $m_2$ is generated from the reconstruction error of pixels, allowing it to outline some image contours while covering adversarial patches. The former may not precisely match the adversarial patch's shape, and the latter might introduce numerous false positives. By combining the two masks, we obtain a final mask that accurately covers the adversarial patch.

\begin{table}[]
\caption{Performance of YOLOv3 with adversarial training on adversarial noises (AT-noise) and adversarial patches (AT-patches).}
    \centering
    \resizebox{0.9\columnwidth}{!}{
    \begin{tabular}{c|c|c|c|c}
        \hline
        \multicolumn{2}{c|}{Attack} & Vanilla & AT-noise & AT-patch \\
        \hline
        \multirow{3}{*}{Targeted HA} & clean & 0.961 & 0.811 & 0.711 \\
         & ~\cite{Thys_2019_CVPR_Workshops}$^*$ & 0.206 & 0.144 & 0.245 \\
         & \cite{Hu_2021_ICCV} & 0.471 & 0.156 & 0.163 \\
        \hline
         \multirow{2}{*}{\shortstack{Untargeted\\HA}} & clean & 0.513 & 0.282 & 0.275 \\
         & ~\cite{lee2019physical} & 0.204 & 0.234 & 0.202  \\
        \hline
        \multirow{2}{*}{AA} & stop sign & 0.806 & 0.929 & 0.750 \\
         & person & 0.952 & 0.778 & 0.618 \\
        \hline
    \end{tabular}}
\label{tab:at evaluation}
\end{table}

\begin{table*}[t]
\caption{Performance ($\text{AP}_{0.5}$) of different methods under AdvPatch~\cite{Thys_2019_CVPR_Workshops} and NaturalPatch~\cite{Hu_2021_ICCV} of different sizes on the INRIA dataset. The gray-background result is obtained from the patch without transformations, while the white-background result is obtained from the patch with transformations such as blurring, rotation, and translation. The values marked in bold are the maximum values. 0.2 (0.25) means the patch height is 0.2 (0.25) times the diagonal of the ground truth bounding box of the object.}
  \centering
  \begin{tabular}{c|Gc|Gc|Gc|cc|Gc|cc}
    \hline
    Model & \multicolumn{2}{c|}{YOLOv2} & \multicolumn{2}{c|}{YOLOv3} & \multicolumn{2}{c|}{YOLOv4} & \multicolumn{2}{c|}{SSD} & \multicolumn{2}{c|}{FRCNN} & \multicolumn{2}{c}{DETR} \\
    \hline
    Clean & \multicolumn{2}{c|}{0.847} & \multicolumn{2}{c|}{0.961} & \multicolumn{2}{c|}{0.955} & \multicolumn{2}{c|}{0.855} & \multicolumn{2}{c|}{0.961} & \multicolumn{2}{c}{0.857} \\
    AdvPatch-0.2 & 0.310 & 0.218 & 0.666 & 0.758 & 0.797 & 0.812 & \cellcolor{gray!30}0.412 & 0.308 & 0.900 & 0.897 & \cellcolor{gray!30}0.872 & 0.860 \\
    AdvPatch-0.25 & 0.136 & 0.059 & 0.206 & 0.341 & 0.417 & 0.473 & \cellcolor{gray!30}0.205 & 0.133 & 0.758 & 0.752 & \cellcolor{gray!30}0.769 & 0.764 \\
    NaturalPatch-0.2 & 0.720 & 0.643 & 0.740 & 0.709 & 0.912 & 0.895 & \multicolumn{2}{c|}{-} & 0.891 & 0.859 & \multicolumn{2}{c}{-} \\
    NaturalPatch-0.25 & 0.637 & 0.476 & 0.471 & 0.448 & 0.877 & 0.810 & \multicolumn{2}{c|}{-} & 0.782 & 0.737 & \multicolumn{2}{c}{-} \\
    \hline
  \end{tabular}
  \label{tab:asr size}
\end{table*}



\begin{figure*}[th]
  \centering
  \begin{subfigure}{0.1\textwidth}
    \centering
    \includegraphics[width=\linewidth]{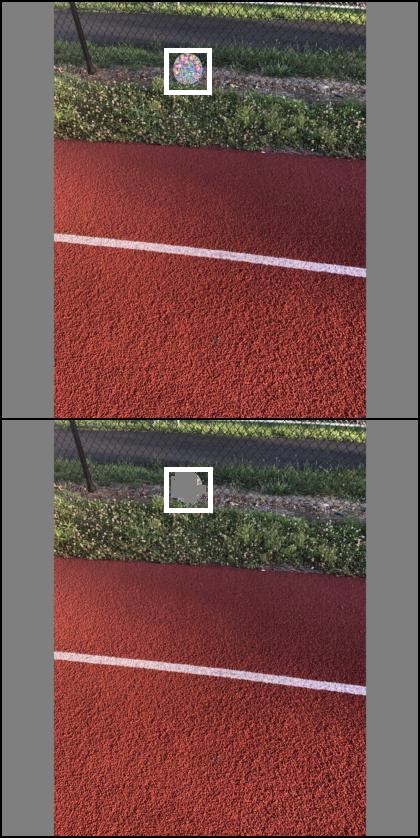}
    \caption{}
    \label{fig:extractor2}
  \end{subfigure}
  \begin{subfigure}{0.1\textwidth}
    \centering
    \includegraphics[width=\linewidth]{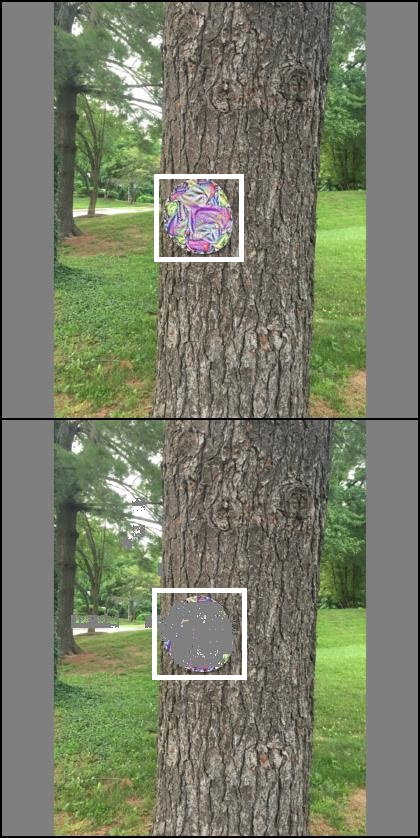}
    \caption{}
    \label{fig:extractor3}
  \end{subfigure}
  \begin{subfigure}{0.1\textwidth}
    \centering
    \includegraphics[width=\linewidth]{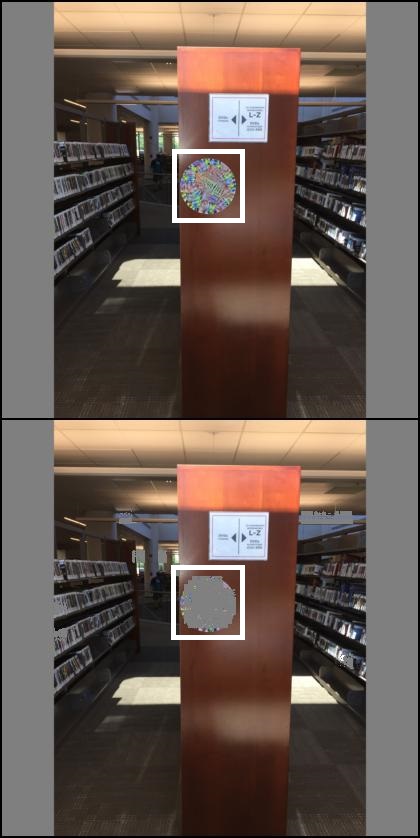}
    \caption{}
    \label{fig:extractor4}
  \end{subfigure}
  \begin{subfigure}{0.1\textwidth}
    \centering
    \includegraphics[width=\linewidth]{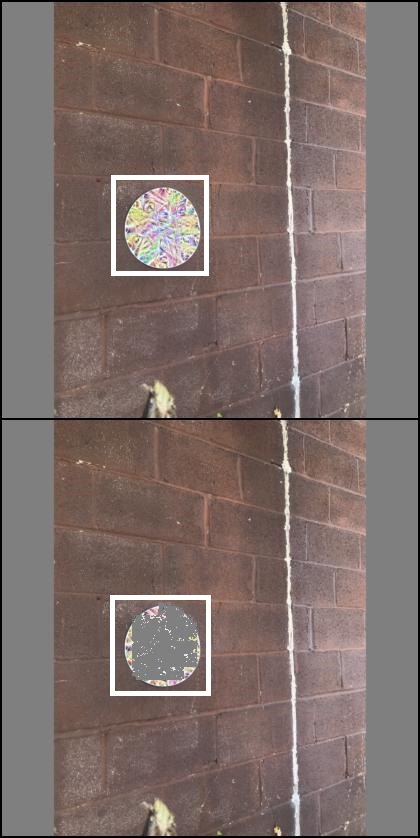}
    \caption{}
    \label{fig:extractor5}
  \end{subfigure}
  \begin{subfigure}{0.1\textwidth}
    \centering
    \includegraphics[width=\linewidth]{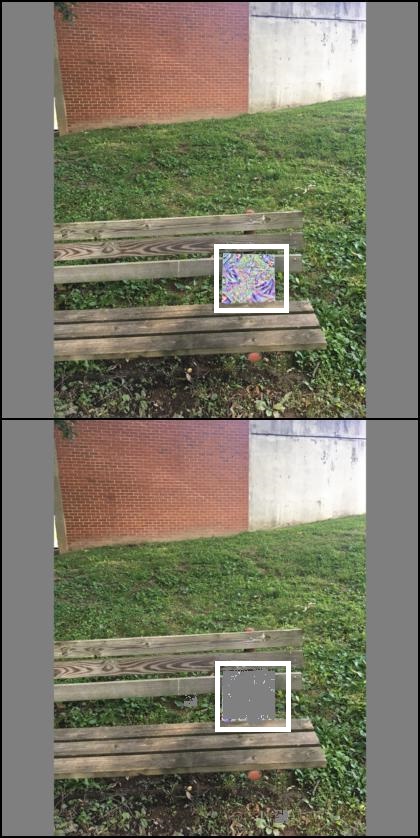}
    \caption{}
    \label{fig:extractor6}
  \end{subfigure}
  \begin{subfigure}{0.1\textwidth}
    \centering
    \includegraphics[width=\linewidth]{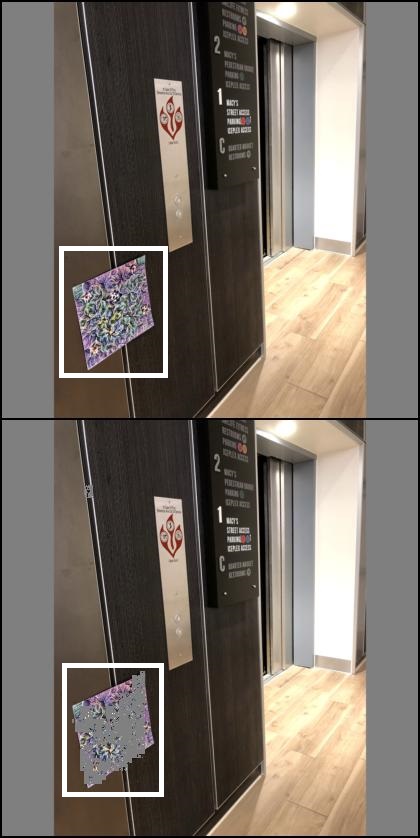}
    \caption{}
    \label{fig:extractor7}
  \end{subfigure}
  \begin{subfigure}{0.1\textwidth}
    \centering
    \includegraphics[width=\linewidth]{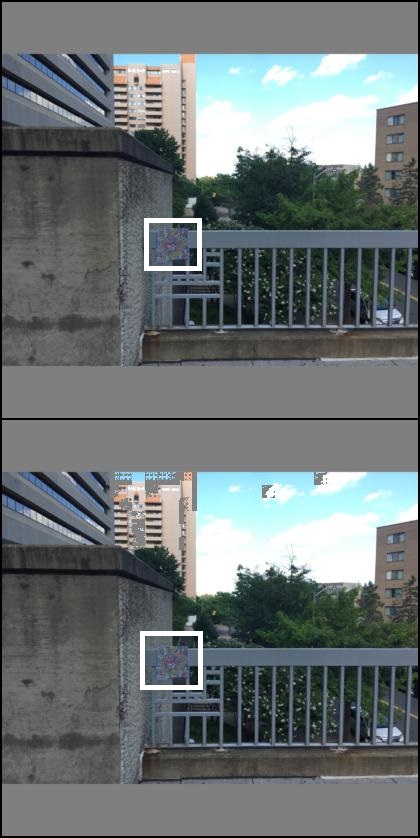}
    \caption{}
    \label{fig:extractor8}
  \end{subfigure}
  \begin{subfigure}{0.1\textwidth}
    \centering
    \includegraphics[width=\linewidth]{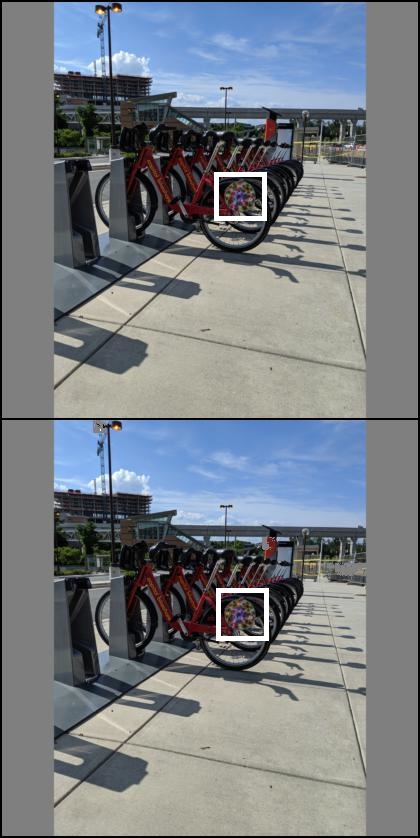}
    \caption{}
    \label{fig:extractor9}
  \end{subfigure}
  \begin{subfigure}{0.1\textwidth}
    \centering
    \includegraphics[width=\linewidth]{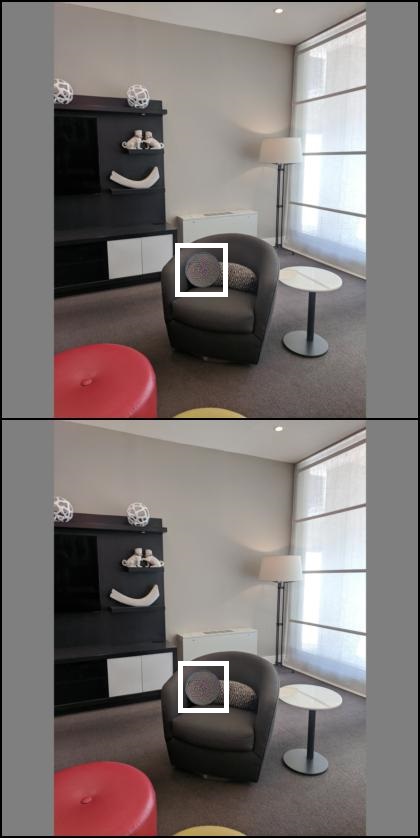}
    \caption{}
    \label{fig:extractor10}
  \end{subfigure}
  \caption{Examples of patch images before and after the distribution extractor and masking process. The upper one is the original image, and the lower one is the masked image. The patch area has been marked by a white bounding box for ease of observation.}
  \label{fig:extractor}
\end{figure*}

\section{Supplementary Evaluation}
\subsection{Evaluation of Offline Defenses}
\label{sec:appendix offline evaluation}
Given that our approach is an online defense, our primary comparisons in Section~\ref{sec:evaluation} involve other online defense methods. However, we also assess the defensive performance of adversarial training, an offline defense strategy. In this evaluation, we train the YOLOv3 detector using adversarial noise and adversarial patches from untargeted Hiding Attacks (HA), testing the detection performance of the patch on different patches, as outlined in Table~\ref{tab:at evaluation}.

The experimental findings indicate that adversarial training significantly impacts the clean performance of the target detector, substantially weakening its detection capability. While adversarial training, utilizing untargeted HA adversarial noise and patches, marginally improves or maintains the model's robustness against PAPatch~\cite{lee2019physical}, it remains ineffective against targeted HA scenarios. Regarding Appearing Attacks (AA), the defense effect of adversarial training exhibits instability. Nevertheless, we posit that the decline in detection accuracy is more a result of weakened model detection capabilities than an enhancement of defense against patches.

\subsection{Evaluation of Targeted HA Patches with Different Sizes}
\label{sec:appendix sizes}
Here, we provide additional details on the configuration of adversarial patches used in our experiments, complementing the information in Section~\ref{sec:exp setup}.
For targeted Hiding Attacks (HA), the height of AdvPatch~\cite{Thys_2019_CVPR_Workshops} is set to 0.2 times the diagonal of the ground truth bounding box for the corresponding target object (i.e., the person) in YOLOv2 and SSD, aligning with the setup in~\cite{Thys_2019_CVPR_Workshops}. For other detectors, including YOLOv3, YOLOv4, Faster RCNN, and DETR, to ensure patch effectiveness, the height is set to 0.25 times the diagonal of the bounding box. Furthermore, the height of NaturalPatch~\cite{Hu_2021_ICCV} is consistently set to 0.25 times the diagonal of the bounding box for all six object detectors.

To more clearly illustrate the rationale behind this setup, we present the attack success rates of patches with different sizes in Table~\ref{tab:asr size}. When the patch height is initially set to 0.2 times the diagonal but does not result in a substantial decrease in detection accuracy (e.g., less than $0.5$), we adjust its size to 0.25 times the diagonal. If this adjustment still fails to significantly impact detection accuracy, we conclude that the adversarial patch is ineffective. The marginal accuracy reduction observed is more likely attributed to the patch's occlusion of the target object itself, and as such, we refrain from further increasing the size.


\begin{figure*}[t]
    \centering
    \begin{subfigure}{0.7\linewidth}
        \centering
        \includegraphics[width=\linewidth]{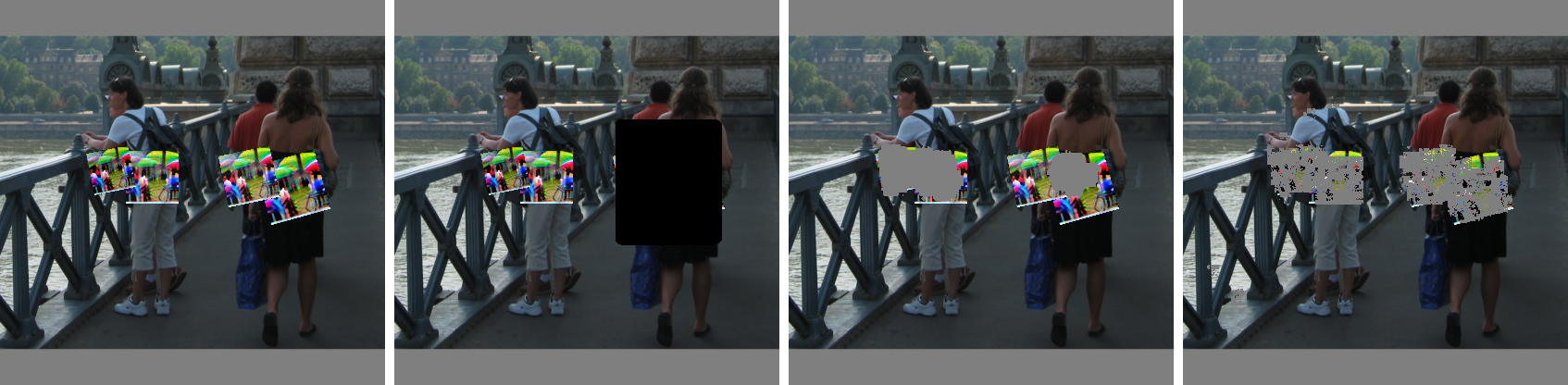}
    \end{subfigure}
    \vspace{3pt}\vfill
    \begin{subfigure}{0.7\linewidth}
        \centering
        \includegraphics[width=\linewidth]{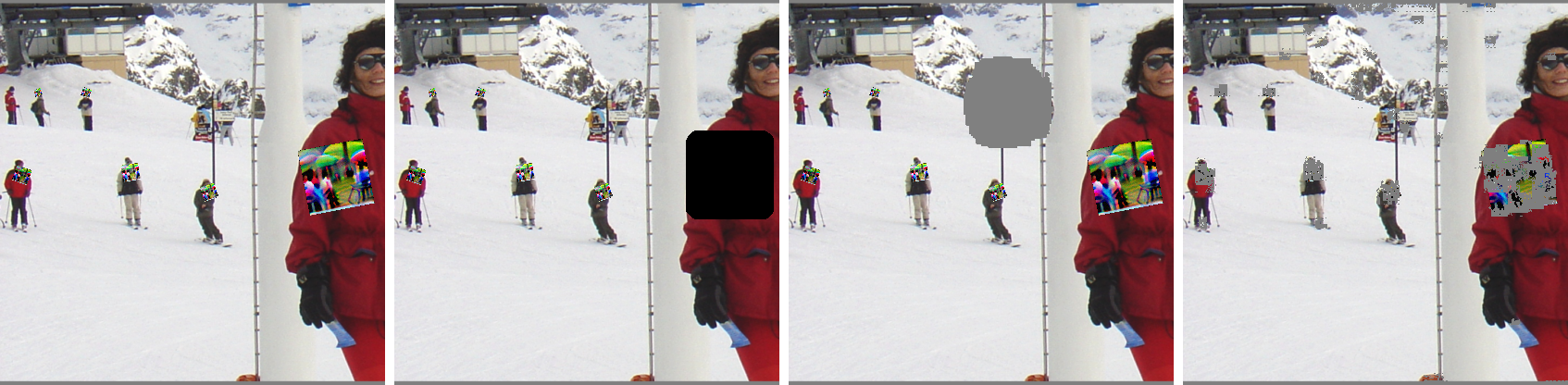}
    \end{subfigure}
    \vspace{3pt}\vfill
    \begin{subfigure}{0.7\linewidth}
        \centering
        \includegraphics[width=\linewidth]{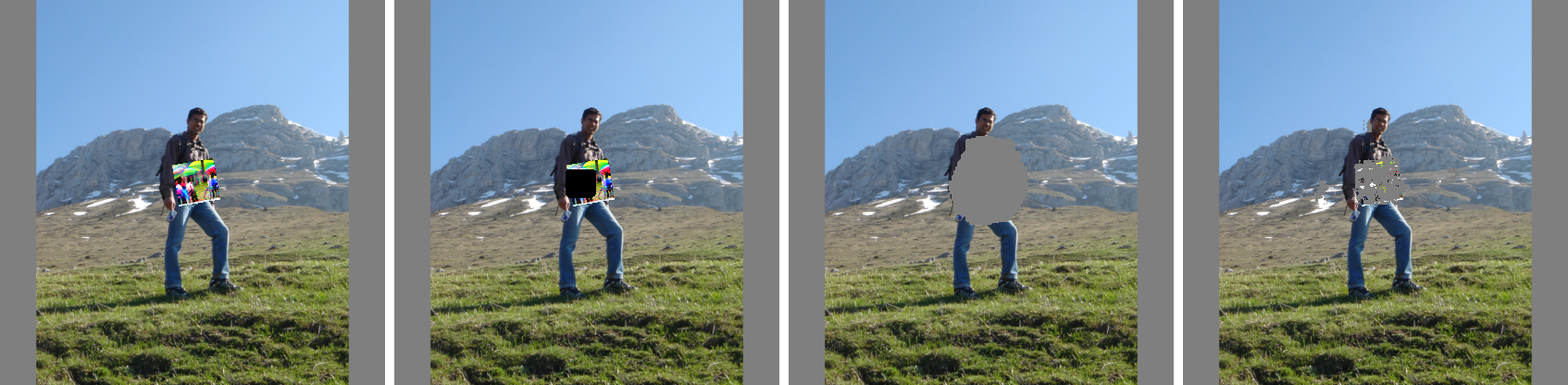}
    \end{subfigure}
    \vspace{3pt}\vfill
    \begin{subfigure}{0.7\linewidth}
        \centering
        \includegraphics[width=\linewidth]{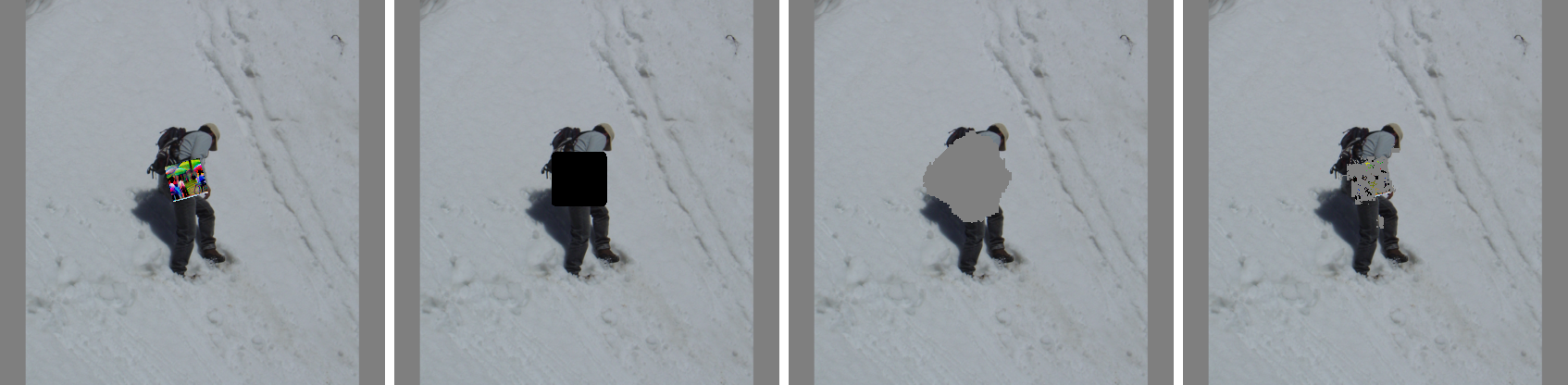}
    \end{subfigure}
    \vspace{3pt}\vfill
    \begin{subfigure}{0.7\linewidth}
        \centering
        \includegraphics[width=\linewidth]{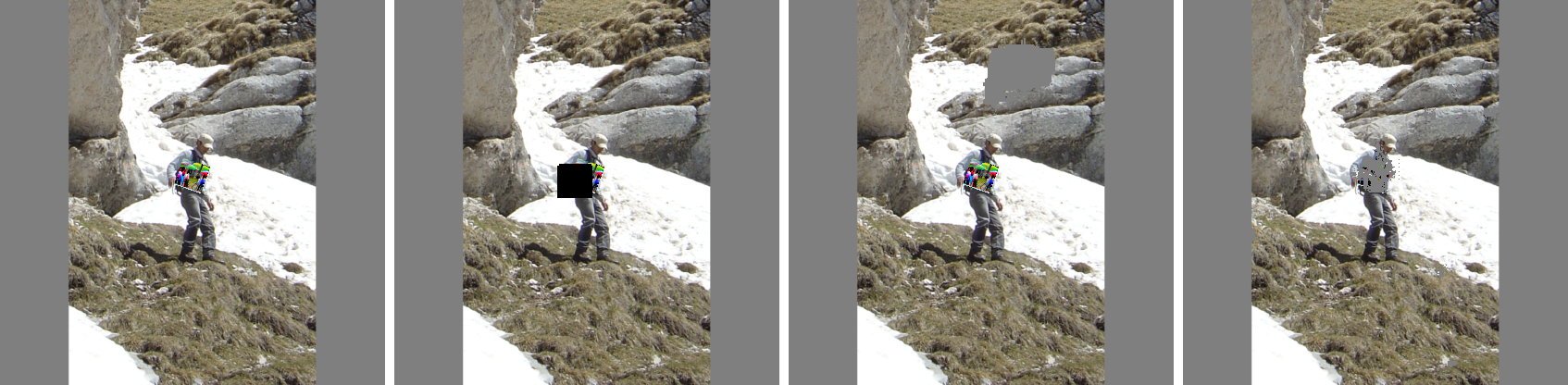}
    \end{subfigure}
    \vspace{3pt}\vfill
    \begin{subfigure}{0.7\linewidth}
        \centering
        \includegraphics[width=\linewidth]{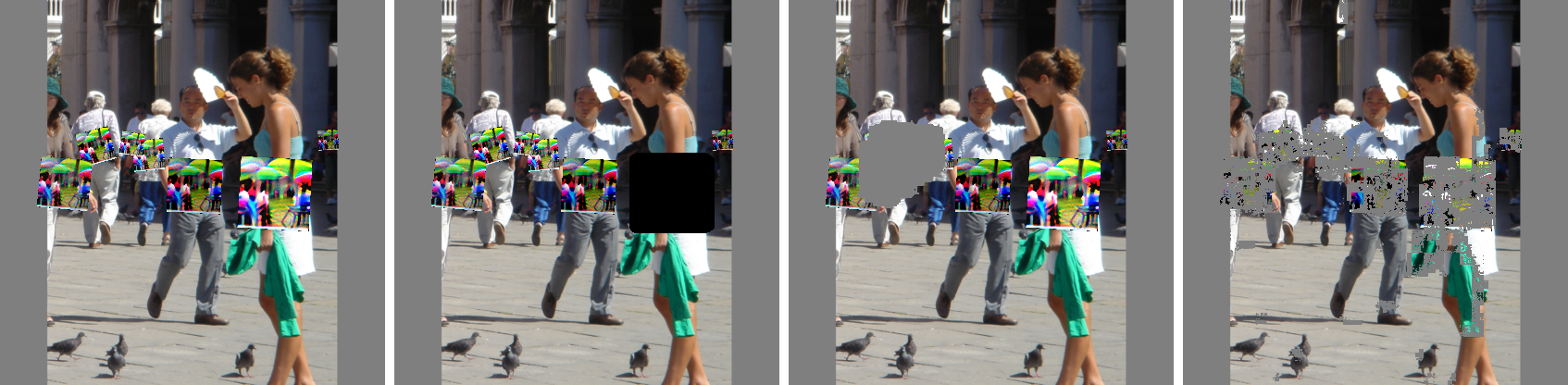}
    \end{subfigure}
    \vspace{1pt}\vfill
    \caption{Examples of the defensive performance of SAC (column 2), Jedi (column 3) and NutNet (column 4). The first column shows the original patched images. Note that SAC uses black masks while Jedi and NutNet use gray ones.}
    \label{fig:more mask examples}
\end{figure*}

\section{More Demonstrations of NutNet}

\subsection{Physical}
\label{sec:appendix more physical}
As an extension to Section~\ref{sec:defense physical}, we introduce more intricate real-world data to further assess the performance of the distribution extractor. In contrast to the evaluation of NutNet's impact on the success rate of printed patches in the physical world outlined in Section~\ref{sec:defense physical}, our focus here is solely on assessing whether the distribution extractor can accurately extract various patches from annotated images across different scenarios.
For this evaluation, we utilize the APRICOT~\cite{braunegg2020apricot} dataset, comprising annotated photographs of printed patches in public locations. We input 873 testing images into NutNet and generate masked images, as depicted in Figure~\ref{fig:extractor}. In each column of images, the upper one represents the original image, while the lower one showcases the masked image.

In well-lit environments, the distribution extractor effectively filters out adversarial patches from the images, as illustrated in Figure~\ref{fig:extractor2}-\ref{fig:extractor7}. We also analyze instances of failure, primarily occurring in scenes with low lighting conditions, as depicted in Figure~\ref{fig:extractor8}-\ref{fig:extractor10}. In such low-light scenarios, the adversarial patch has largely lost its efficacy, making it challenging even for humans to promptly identify. Despite these occasional failures, we consider them acceptable. It's important to note that capturing images of patches in the physical world poses additional challenges for extracting distributions compared to the digital world. Nevertheless, the distribution extractor demonstrates its effectiveness in filtering out adversarial patches.



\subsection{Digital}
\label{sec:appendix more digital}
As an extension of Figure~\ref{fig:mask examples} in Section~\ref{sec:robustness}, additional examples showcasing the masking performance of our defense, along with Jedi and SAC, are provided in Figure~\ref{fig:more mask examples}. The second-row example serves as a notable representation of the capabilities of the three defense methods. SAC primarily masks larger adversarial patches, while Jedi tends to mask regions of the background with higher entropy rather than the actual adversarial patches. In contrast, NutNet demonstrates precise detection and masking of all adversarial patches, irrespective of their size or location.

\end{document}